\documentclass[aps,prl,preprint,superscriptaddress,citeautoscript,floatfix]{revtex4-2}
\usepackage{amsmath}
\usepackage{amssymb}

\usepackage{graphicx}
\usepackage{epstopdf}
\usepackage{hyperref}

\usepackage{pdfpages}
\makeatletter
\AtBeginDocument{\let\LS@rot\@undefined}
\makeatother

\usepackage{color,soul}

\setstcolor{red}
\usepackage[normalem]{ulem}
\usepackage{xcolor}

\let\oldhat\hat
\renewcommand{\vec}[1]{\mathbf{#1}}
\renewcommand{\hat}[1]{\oldhat{\mathbf{#1}}}

\begin{document}


\title{Anomalous normal fluid response in a chiral superconductor UTe$_2$}

\author{Seokjin Bae}\email{sjbae@terpmail.umd.edu, anlage@umd.edu}
\affiliation{Maryland Quantum Materials Center, Department of Physics, University of Maryland, College Park, MD 20742, USA}
\affiliation{Current address: Materials Research Laboratory, University of Illinois Urbana-Champaign, Urbana, Illinois 61801, USA}

\author{Hyunsoo Kim}
\affiliation{Maryland Quantum Materials Center, Department of Physics, University of Maryland, College Park, MD 20742, USA}
\affiliation{Current address: Department of Physics and Astronomy, Texas Tech University, Lubbock, Texas 79410, USA}

\author{Yun Suk Eo}
\affiliation{Maryland Quantum Materials Center, Department of Physics, University of Maryland, College Park, MD 20742, USA}

\author{Sheng Ran}
\affiliation{Maryland Quantum Materials Center, Department of Physics, University of Maryland, College Park, MD 20742, USA}
\affiliation{NIST Center for Neutron Research, National Institute of Standards and Technology, Gaithersburg, MD 20899, USA}
\affiliation{Current address: Department of Physics, Washington University, St. Louis, MO 63130, USA}

\author{I-lin Liu}
\affiliation{Maryland Quantum Materials Center, Department of Physics, University of Maryland, College Park, MD 20742, USA}
\affiliation{NIST Center for Neutron Research, National Institute of Standards and Technology, Gaithersburg, MD 20899, USA}

\author{Wesley T. Fuhrman}
\affiliation{Maryland Quantum Materials Center, Department of Physics, University of Maryland, College Park, MD 20742, USA}

\author{Johnpierre Paglione}
\affiliation{Maryland Quantum Materials Center, Department of Physics, University of Maryland, College Park, MD 20742, USA}
\affiliation{The Canadian Institute for Advanced Research, Toronto, Ontario M5G 1Z8, Canada}

\author{Nicholas P. Butch}
\affiliation{Maryland Quantum Materials Center, Department of Physics, University of Maryland, College Park, MD 20742, USA}
\affiliation{NIST Center for Neutron Research, National Institute of Standards and Technology, Gaithersburg, MD 20899, USA}

\author{Steven M. Anlage}\email{sjbae@terpmail.umd.edu, anlage@umd.edu}
\affiliation{Maryland Quantum Materials Center, Department of Physics, University of Maryland, College Park, MD 20742, USA}

\date{\today}

\begin{abstract}
Chiral superconductors have been proposed as one pathway to realize Majorana normal fluid at its boundary. However, the long-sought 2D and 3D chiral superconductors with edge and surface Majorana normal fluid are yet to be conclusively found. Here we report evidence for chiral spin-triplet pairing state of UTe2 with surface normal fluid response. The microwave surface impedance of the UTe$_2$ crystal was measured and converted to complex conductivity, which is sensitive to both normal and superfluid responses. The anomalous residual normal fluid conductivity supports the presence of a significant normal fluid response. The superfluid conductivity follows the temperature behavior predicted for axial spin-triplet state, which is further narrowed down to chiral spin-triplet state with evidence of broken time-reversal symmetry. Further analysis excludes trivial origins for the observed normal fluid response. Our findings suggest that UTe$_2$ can be a new platform to study exotic topological excitations in higher dimension.
\end{abstract}



\maketitle

\section{Introduction}
\par Topological insulators, with non-zero topological invariants, possess metallic states at their boundary \cite{Hasan2010RMP}. Chiral superconductors, a type of topological superconductors with non-zero topological invariants, possess Majorana fermions at their boundary \cite{Schnyder2008PRB, Kallin2016RPP, Kozii2016SciAdv, Sato2017RPP}. Majorana fermions are an essential ingredient to establish topological quantum computation \cite{Kitaev2003AnnPhy}. Hence great effort has been given to search for chiral superconducting systems. So far, evidence for the 1D example has been found from a semiconductor nanowire with end-point Majorana states \cite{Deng2016Science}. However, 2D and 3D chiral superconductors with a surface Majorana normal fluid has not been unequivocally found \cite{Kallin2016RPP}. Recently, a newly discovered heavy-fermion superconductor UTe$_2$ \cite{SRan2019Science} is proposed to be a long-sought 3D chiral superconductor with evidence of the chiral in-gap state from a scanning tunneling microscopy (STM) study \cite{LJiao2020Nature}. This raises a great deal of interest in the physics community to independently establish the existence of the normal fluid response, determining whether or not the response is intrinsic, and identifying the nature of the pairing state of UTe$_2$.

\par To address these three questions, the microwave surface impedance of a UTe$_2$ crystal was measured by the dielectric resonator technique (See Supplementary Note 1, 2, and 3). The obtained impedance was converted to the complex conductivity, where the real part is sensitive to the normal fluid response and the imaginary part is sensitive to the superfluid response. By examining these results, here we confirm the existence of the significant normal fluid response of UTe$_2$, verify that the response is intrinsic, and identify that the gap structure is consistent with the chiral spin-triplet pairing state.

\section{Results}
\par
\begin{figure}
		\includegraphics[width=1\columnwidth]{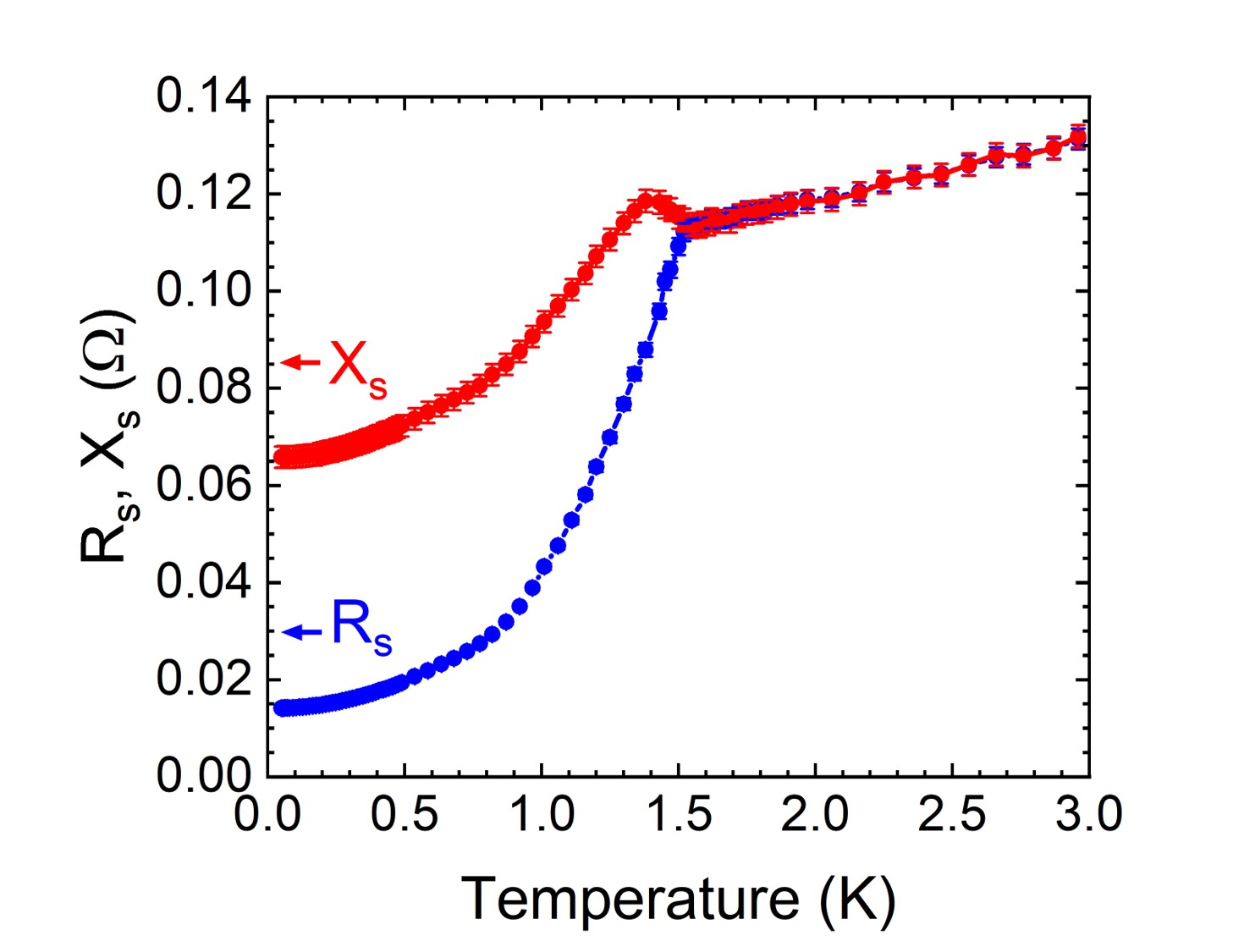}
		\caption{\label{fig:Fig1} \textbf{Microwave surface impedance of a UTe$_2$ single crystal.} The measured temperature dependence of the surface impedance of a UTe$_2$ sample at 11.26 GHz. The blue curve represents the surface resistance $R_s$ and the red curve represents the surface reactance $X_s$.}
\end{figure}

 \textbf{Anomalous residual normal fluid response.} Figure \ref{fig:Fig1} shows the surface impedance $Z_s = R_s + iX_s$ of the sample as a function of temperature. The surface resistance $R_s$ decreases monotonically below $\approx 1.55$ K and reaches a surprisingly high residual value $R_s(0)\approx 14$ m$\Omega$ at 11.26 GHz. This value is larger by an order of magnitude than that of another heavy-fermion superconductor CeCoIn$_5$ ($R_s(0)\approx 0.9$ m$\Omega$ at 12.28 GHz) \cite{Truncik2013NatComm}. Subsequently its electrical resistance was determined by transport measurement and a mid-point $T_c$ of 1.57 K was found.
 
\par
\begin{figure}
		\includegraphics[width=1\columnwidth]{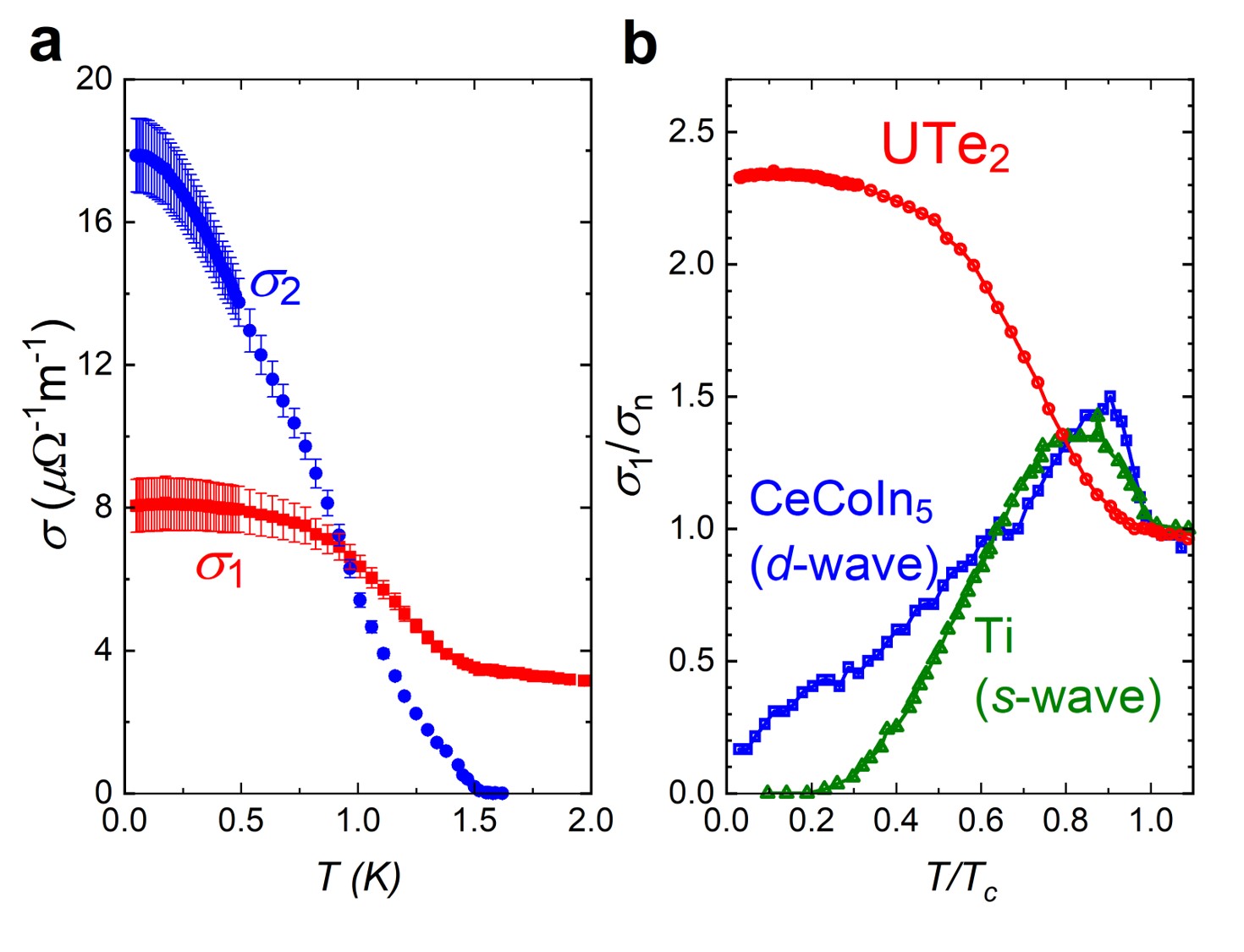}
		\caption{\label{fig:Fig2} \textbf{Anomalous residual conductivity of UTe$_2$ compared to superconductors with other pairing states.} (a) Real (red) and imaginary (blue) part of the complex conductivity $\tilde{\sigma}=\sigma_1 - i\sigma_2$ of the UTe$_2$ sample at 11.26 GHz. (b) Normalized (by $\sigma_n=\sigma_1(T_c)$) real part of conductivity of UTe$_2$ (red), a line nodal $d_{x^2-y^2}$-wave superconductor CeCoIn$_5$ (blue) \cite{Truncik2013NatComm}, and a fully gapped $s$-wave superconductor Ti (green) \cite{Thiemann2018PRB} versus reduced temperature $T/T_c$. All measurements are done with the same, low frequency-to-gap ratio of $\hbar\omega/2\Delta_0\approx 0.08$.}
\end{figure}

 With the surface impedance, the complex conductivity $\tilde{\sigma}=\sigma_1 - i\sigma_2$ of the sample can be calculated. In the local electrodynamics regime (Supplementary Note 4), one has $Z_s = \sqrt{i\mu_0 \omega/\tilde{\sigma}}$. Figure \ref{fig:Fig2}(a) shows $\sigma_1$ and $\sigma_2$ of the sample as a function of temperature. Here, an anomalous feature is the monotonic increase of $\sigma_1(T)$ as $T$ decreases below $T_c$. Note that $\sigma_1$ is a property solely of the normal fluid. For superconductors with a topologically trivial pairing state, most of the normal fluid turns into superfluid and is depleted as $T\rightarrow 0$. As a result, in the low temperature regime, $\sigma_1$ shows a strong decrease as temperature decreases, and is expected to reach a theoretically predicted residual value $\sigma_1(0)/\sigma_1(T_c)=0$ (for fully gapped $s$-wave \cite{Klein1994PRB}), $< 0.1$ (for the bulk state of a point nodal $p$-wave \cite{Hirschfeld1989PRB}) and $0.1 \sim 0.3$ (for line nodal $d_{x^2-y^2}$-wave \cite{Hirschfeld1993PRL,KZhang1994PRL}). As shown in Fig. \ref{fig:Fig2}(b), this behavior is observed for the case of Ti \cite{Thiemann2018PRB} ($s$-wave) as well as CeCoIn$_5$ \cite{Truncik2013NatComm} ($d_{x^2-y^2}$-wave). In contrast, the UTe$_2$ crystal shows a monotonic increase in $\sigma_1$ as the temperature decreases and reaches a much larger $\sigma_1(0)/\sigma_1(T_c)=2.3$, implying the normal fluid conduction channel is still active and provides a significant contribution even at the lowest temperature.

\par \textbf{Axial triplet pairing state from superfluid response.} Another property one can extract from the complex conductivity is the effective penetration depth. The imaginary part of the complex conductivity $\sigma_2(T)$ determines the absolute value of the effective penetration depth at each temperature as $\sigma_2(T) = 1/\mu_0\omega\lambda^2_\textrm{eff}(T)$. Once the absolute value of the penetration depth is known, the normalized superfluid density can be calculated as $\rho_s(T) = \lambda^2_\textrm{eff}(0)/\lambda^2_\textrm{eff}(T)$ (See Methods for how $\lambda_\textrm{eff}(0)$ is determined), and its low temperature behavior is determined by the low energy excitations of the superconductor, which is sensitive to the pairing state \cite{Prozorov2006SST}. The $s$ and $d$-wave pairing states, representative spin-singlet pairing states, are inconsistent with our penetration depth data (See Supplementary Note 5). More crucially, singlet states cannot explain either the reported upper critical field $H_{c2}$ which is larger than the paramagnetic limiting field \cite{SRan2019Science}, or the absence of a change in the Knight shift across the $T_c$ \cite{SRan2019Science,Nakamine2019JPSJ}. Thus, only the spin-triplet pairing states are discussed below.

\par
\begin{figure}
		\includegraphics[width=0.7\columnwidth]{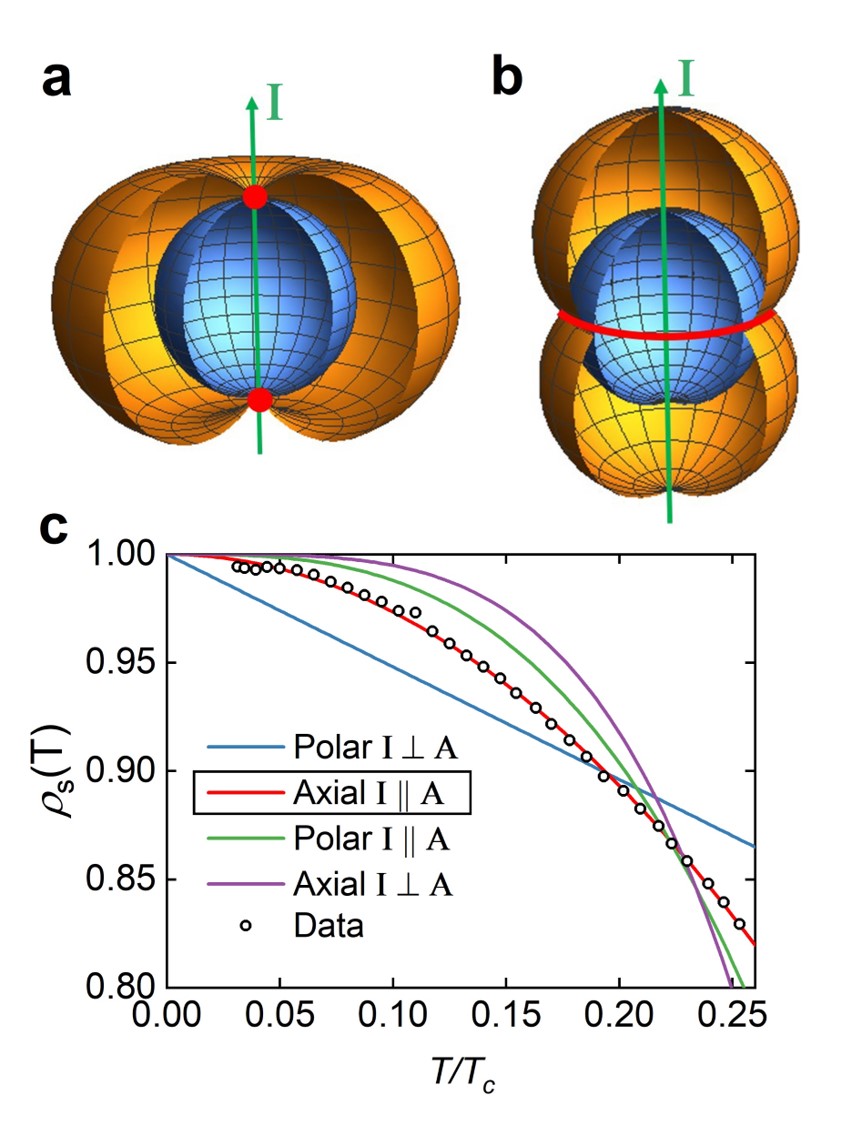}
		\caption{\label{fig:Fig4} \textbf{Evidence for the axial triplet pairing state from the superfluid density.} (a) A schematic plot of the gap magnitude $|\Delta(\vec{k})|$ (orange) and the Fermi surface (blue) in momentum space for the axial triplet pairing state. $\vec{I}$ represents the symmetry axis of the gap function. Note that two point nodes (red) exist along the symmetry axis. (b) For the case of the polar state. A line node (red) exists along the equatorial plane. (c) Low temperature behavior of the normalized superfluid density $\rho_s(T)$ in UTe$_2$ with best fits for various triplet pairing states, and relative direction between the symmetry axis $\vec{I}$ and the vector potential $\vec{A}$. Since $\vec{I}$ connects the two point nodes of the gap of the axial pairing state and the measurement surveys the $ab$-plane electrodynamics, one can conclude that the point nodes are located near the $ab$-plane. Evidence of broken time-reversal symmetry \cite{Ian2020arXiv} further narrows down the pairing state from axial to chiral.}
\end{figure}

 For a spin-triplet pairing state, $\rho_s(T)$ follows different theoretical low temperature behaviors depending on two factors \cite{Gross1986ZPhyB,Prozorov2006SST}. One is whether the magnitude of the energy gap $|\Delta(\hat{\vec{k}},T)|$ follows that of an axial state $\Delta_0(T)|\hat{\vec{k}}\times \hat{\vec{I}}|$ (Fig. \ref{fig:Fig4}(a)) or a polar state  $\Delta_0(T)|\hat{\vec{k}}\cdot \hat{\vec{I}}|$ (Fig. \ref{fig:Fig4}(b)), where $\Delta_0(T)$ is the gap maximum. The other is whether the vector potential direction $\hat{\vec{A}}$ is parallel or perpendicular to the symmetry axis of the gap $\hat{\vec{I}}$. Figure \ref{fig:Fig4}(c) shows fits of $\rho_s(T)$ to the theoretical behavior (Ref. \cite{Gross1986ZPhyB,Prozorov2006SST} and Supplementary Note 8) of the various triplet pairing states. Apparently, the data follows the behavior of the axial pairing state with the direction of the current aligned to $\hat{\vec{I}}$. The axial state can be either chiral or helical depending on the presence or absence of time-reversal symmetry breaking (TRSB) of the system \cite{Kallin2016RPP}. Recently, direct evidence for TRSB in this system was found by a finite polar Kerr rotation angle developing below $T_c$ \cite{Ian2020arXiv}. Also, a specific heat study, a sensitive probe to resolve multiple superconducting transitions, showed two jumps near 1.6 K \cite{Ian2020arXiv}. This implies that two nearly degenerate order parameters coexist, allowing the chiral pairing state from the group theory perspective \cite{Ian2020arXiv}. Therefore, one can argue that UTe$_2$ shows $\rho_s(T)$ consistent with the chiral triplet pairing state. In addition, since the symmetry axis connects the two point nodes of the chiral pairing order parameter, and the measurement surveys the $ab$-plane electrodynamics, one can further conclude that the point nodes are located near the $ab$-plane. The low temperature asymptote of $\rho_s(T)$ in this case is given as $\rho_s(T)=1-\pi^2(k_BT/\Delta_0(0))^2$. The fitting in Fig. \ref{fig:Fig4}(c) yields an estimate for the gap size $\Delta_0(0)=1.923 \pm 0.002  k_BT_c \approx 0.265$ meV.  Note that a recent STM study \cite{LJiao2020Nature} measures a similar gap size (0.25 meV).

\par \textbf{Examination of extrinsic origins.} Our study shows evidence for the chiral triplet pairing state and a substantial amount of normal fluid response in the ground state of UTe$_2$. Before attributing this residual normal fluid response to an intrinsic origin, one must first examine the possibilities of an extrinsic origin. One of the possible extrinsic origins would be a large bulk impurity scattering rate $\Gamma_\textrm{imp}$. However, if one fits the temperature dependence of the normalized superfluid density to $\rho_s(T)=1-aT^2$ and compares the estimated coefficient $a$ with the modified theoretical asymptote for the chiral triplet state which includes $\Gamma_\textrm{imp}$ \cite{Gross1986ZPhyB},
\begin{equation}
\rho_s(T) = 1 - \frac{1}{1-3\frac{\Gamma_\textrm{imp}}{\Delta_0(0)}\left(\frac{\pi}{2}\ln2 - 1 \right)}\frac{\pi^2}{1-\frac{\pi\Gamma_\textrm{imp} }{2\Delta_0(0)}}\left( \frac{k_BT}{\Delta_0(0)} \right)^2  ,
\end{equation}
one obtains a quadratic equation for $(\Gamma_\textrm{imp}, \Delta_0(0))$, and two solutions: $\Gamma_\textrm{imp} \leq 0.06\Delta_0(0)$ and 
$\Gamma_\textrm{imp} \geq 4.36\Delta_0(0)$ for the range of $\Delta_0(0)\leq 0.280$ meV. For a nodal superconductor, the large $\Gamma_\textrm{imp} \geq 4.36\Delta_0(0)$ suppresses $T_c$ more than 80 \% \cite{Radtke1993PRB,Mackenzie1998PRL}. Such suppression was not observed for the sample in this study. This leaves the vanishingly small $\Gamma_\textrm{imp}$ as the only physically reasonable choice. Note that this small bulk $\Gamma_\textrm{imp}$ is consistent with the absence of the residual linear term in the thermal conductivity \cite{Tristin2019PRB}, both implying the clean limit $\Gamma_\textrm{imp} \ll 2\Delta_0(0)$. These results are inconsistent with the impurity-induced bulk normal fluid scenario. Note that the temperature independent $\Gamma_\textrm{imp}$ here is much different from the scattering rate above $T_c$ (See Supplementary Note 4), possibly suggesting the presence of highly temperature dependent inelastic scattering due to spin fluctuations in the normal state.

\par Another possibility is a quasiparticle response excited by the microwave photons of the measurement signal. However, this scenario is also improbable because the maximum energy gap $\Delta_0(0) = 265$ $\mu$eV is much larger than that of the microwave photon $E_\textrm{ph}=45$ $\mu$eV used here. At this low $E_\textrm{ph}/\Delta_0(0)$ ratio, even when the upper limit of $\Gamma_\textrm{imp}=0.06\Delta_0(0)\sim0.1 k_BT_c$ is assumped, a theoretical estimate \cite{Hirschfeld1989PRB} predicts only a small residual normal response from the bulk states $\sigma_1(0)/\sigma_n < 0.1$, which cannot explain the measured value of $\sigma_1(0)/\sigma_n = 2.3$.

\par \textbf{Possible intrinsic origin.} With several candidates for extrinsic origin excluded, now one can consider the possibility of an intrinsic origin. One important constraint to consider is the recent bulk thermal conductivity measurement in UTe$_2$ \cite{Tristin2019PRB}. It revealed the absence of a residual linear term as a function of temperature in the thermal conductivity, implying the absence of residual normal carriers, at least in the bulk (See Supplementary Note 7). This result suggests the microwave conductivity data can be explained by a combination of a surface normal fluid and bulk superfluid. Topologically trivial origins for the surface normal fluid (e.g. pair breaking rough surface) are first examined and shown to be inconsistent with the monotonic increase of $\sigma_1$ as $T\rightarrow 0$ in Fig. \ref{fig:Fig2}(b) (See supplementary Note 11). Instead, considering the evidence of the chiral triplet pairing state from the superfluid density analysis and polar Kerr rotation measurement \cite{Ian2020arXiv}, a more pluasible source of the surface normal fluid is the gapless chiral-dispersing surface states of a 3D chiral superconductor \cite{Kozii2016SciAdv}. Point nodes of a chiral superconductor can possess non-zero Chern number with opposite sign. These point nodes in the superconducting gap are analogous to the Weyl points in the bulk energy bands of a Weyl semimetal. The nodes are predicted to introduce gapless surface Majorana arc states which connect them \cite{Sato2017RPP}. Evidence for these surface states in UTe$_2$ is seen in a chiral in-gap density of states from an STM study \cite{LJiao2020Nature}.

\par If this scenario is true, the anomalous monotonic increase in $\sigma_1$ down to zero temperature from this surface impedance study (Fig. \ref{fig:Fig2}(b)) may be understood as the enhancement of the scattering lifetime of the surface normal fluid. With its chiral energy dispersion, direct backscattering can be suppressed. As the temperature decreases, the superconducting gap $\Delta_0(T)$ which provides this topological protection increases, while thermal fluctuations $k_BT$ which poison the protection decreases. As a result, the suppression of the backscattering becomes stronger as $T\rightarrow 0$, which can end up enhancing the surface scattering lifetime (and $\sigma_1$). Although this argument is speculative at the moment, we hope the anomalous behavior of the $\sigma_1(T)$ reported in this work motivates quantitative theoretical investigation for the microwave response of the topological surface state of chiral superconductors.

\par In conclusion, our findings imply that UTe$_2$ may be the first example of a 3D chiral spin-triplet superconductor with a surface Majorana normal fluid. With topological excitations in a higher dimension, this material can be a new platform to pursue unconventional superconducting physics, and act as a setting for topological quantum computation.
 
\newpage 
\section*{METHODS}
\par \textbf{Growth and preparation of UTe$_2$ single crystals.} The single crystal sample of UTe$_2$ was grown by the chemical vapor transport method using iodine as the transport agent \cite{SRan2019Science}. For the microwave surface impedance measurement, the top and bottom $ab$-plane facets were polished on a 0.5 $\mu$m alumina polishing paper inside a nitrogen-filled glove bag (O$_2$ content $<$ 0.04 \%). After polishing was done, the sample was encapsulated by Apeizon N-grease (See Supplementary Note 10) before being taken out from the bag, and then mounted to the resonator so that the sample is protected from oxidization. Long-term storage of the sample is done in a glove box with O$_2$ content $<$ 0.5 ppm. Note that the electrical properties of oxidized uranium are summarized in Supplementary Note 9. The sample size after polishing is about 1.5$\times$0.7$\times$0.3 mm$^3$ with the shortest dimension being along the crystallographic $c$-axis of the orthorhombic structure. The mid-point $T_c$ of the sample from DC transport measurements was 1.57 K.

\par \textbf{Microwave surface impedance measurement.} The measurement setup, data processing procedure, and interpretation are decribed in detail in the Supplementary Note 1 and 2.

\par \textbf{Determining of the value of the zero temperature absolute penetration depth and comparison to other Uranium based superconductors.} The effective penetration depth at each temperature ($T\geq 50$ mK) can be obtained from $\sigma_2(T) = 1/\mu_0\omega\lambda^2_\textrm{eff}(T)$, where $\sigma_2(T)$ is obtained by the surface impedance $Z_s(T)$ data. The effective penetration depth at zero temperature can be obtained by extrapolating the data with a power law fit $\lambda_\textrm{eff}(T)-\lambda_\textrm{eff}(0)=aT^c$ over the low temperature regime $T<0.3T_c$, resulting in $\lambda_\textrm{eff}(0)=791$ nm and $c=2.11$. This value is similar to those found in the Uranium based ferromagnetic superconductor series such as UCoGe ($\lambda_\textrm{eff}(0)\sim 1200$ nm) \cite{Huy2008Thesis} and URhGe ($\lambda_\textrm{eff}(0)\sim 900$ nm) \cite{Aoki2001Nat}, where UTe$_2$ represents the paramagnetic end member of the series \cite{SRan2019Science}. This result is also consistent with recent muon-spin rotation measurements on UTe$_2$ which concluded $\lambda_\textrm{eff}(0)\sim 1000$ nm \cite{Sundar2019PRB}.

\par \textbf{Error bar of the fitting parameters of the normalized superfluid density.} In the fitting of the normalized superfluid density, the error bar of the fitting parameter (e.g., $\Delta_0$) was determined by the deviation from the estimated value which increases the root-mean-square error of the fit by 1\%.

\section*{Acknowledgements}
\par This work is supported by NSF grant No. DMR - 1410712, DOE grants No. DE-SC 0017931 (support of S.B.), DE-SC 0018788 (measurements), DE-SC-0019154 (measurements), Gordon and Betty Moore Foundation's EPiQS Initiative through Grant GBMF9071 (synthesis), and the Maryland Quantum Materials Center.

\section*{Data availability} The datasets generated and analysed in this work are available from the corresponding author on reasonable request. Source data for Figs. 1–3 and Supplementary Fig. 4 are provided with the paper.

\section*{Competing interests} The authors declare no competing interests

\section*{AUTHOR CONTRIBUTION}
\par S. B., N. P. B., and S. M. A. conceived the project. S. B. polished the sample, conducted the microwave surface impedance measurements, and carried out analysis on the complex conductivity. H. K. and Y. S. E. performed the transport measurements. S. R., I. L., and W. T. F. grew the UTe$_2$ single crystal used in this study. S. B., H. K., J. P., N. P. B., and S. M. A. interpreted the results. S. B. and S. M. A. wrote the manuscript with input from other authors.

\section*{Refernces}
\bibliography{Ref_201908}

\begin{thebibliography}{10}
\expandafter\ifx\csname url\endcsname\relax
  \def\url#1{\texttt{#1}}\fi
\expandafter\ifx\csname urlprefix\endcsname\relax\def\urlprefix{URL }\fi
\providecommand{\bibinfo}[2]{#2}
\providecommand{\eprint}[2][]{\url{#2}}

\bibitem{Hasan2010RMP}
\bibinfo{author}{Hasan, M.~Z.} \& \bibinfo{author}{Kane, C.~L.}
\newblock \bibinfo{title}{Colloquium: Topological insulators}.
\newblock \emph{\bibinfo{journal}{Rev. Mod. Phys.}}
  \textbf{\bibinfo{volume}{82}}, \bibinfo{pages}{3045--3067}
  (\bibinfo{year}{2010}).
\newblock \urlprefix\url{https://link.aps.org/doi/10.1103/RevModPhys.82.3045}.

\bibitem{Schnyder2008PRB}
\bibinfo{author}{Schnyder, A.~P.}, \bibinfo{author}{Ryu, S.},
  \bibinfo{author}{Furusaki, A.} \& \bibinfo{author}{Ludwig, A. W.~W.}
\newblock \bibinfo{title}{Classification of topological insulators and
  superconductors in three spatial dimensions}.
\newblock \emph{\bibinfo{journal}{Phys. Rev. B}} \textbf{\bibinfo{volume}{78}},
  \bibinfo{pages}{195125} (\bibinfo{year}{2008}).
\newblock \urlprefix\url{https://link.aps.org/doi/10.1103/PhysRevB.78.195125}.

\bibitem{Kallin2016RPP}
\bibinfo{author}{Kallin, C.} \& \bibinfo{author}{Berlinsky, J.}
\newblock \bibinfo{title}{Chiral superconductors}.
\newblock \emph{\bibinfo{journal}{Rep. Prog. Phys.}}
  \textbf{\bibinfo{volume}{79}}, \bibinfo{pages}{054502}
  (\bibinfo{year}{2016}).
\newblock
  \urlprefix\url{https://doi.org/10.1088%2F0034-4885%2F79%2F5%2F054502}.

\bibitem{Kozii2016SciAdv}
\bibinfo{author}{Kozii, V.}, \bibinfo{author}{Venderbos, J. W.~F.} \&
  \bibinfo{author}{Fu, L.}
\newblock \bibinfo{title}{Three-dimensional majorana fermions in chiral
  superconductors}.
\newblock \emph{\bibinfo{journal}{Sci. Adv.}} \textbf{\bibinfo{volume}{2}},
  \bibinfo{pages}{e1601835} (\bibinfo{year}{2016}).
\newblock
  \urlprefix\url{https://advances.sciencemag.org/content/2/12/e1601835}.

\bibitem{Sato2017RPP}
\bibinfo{author}{Sato, M.} \& \bibinfo{author}{Ando, Y.}
\newblock \bibinfo{title}{Topological superconductors: a review}.
\newblock \emph{\bibinfo{journal}{Rep. Prog. Phys.}}
  \textbf{\bibinfo{volume}{80}}, \bibinfo{pages}{076501}
  (\bibinfo{year}{2017}).
\newblock \urlprefix\url{https://doi.org/10.1088%2F1361-6633%2Faa6ac7}.

\bibitem{Kitaev2003AnnPhy}
\bibinfo{author}{Kitaev, A.}
\newblock \bibinfo{title}{Fault-tolerant quantum computation by anyons}.
\newblock \emph{\bibinfo{journal}{Ann. Phys.}} \textbf{\bibinfo{volume}{303}},
  \bibinfo{pages}{2 -- 30} (\bibinfo{year}{2003}).
\newblock
  \urlprefix\url{http://www.sciencedirect.com/science/article/pii/S0003491602000180}.

\bibitem{Deng2016Science}
\bibinfo{author}{Deng, M.~T.} \emph{et~al.}
\newblock \bibinfo{title}{Majorana bound state in a coupled quantum-dot
  hybrid-nanowire system}.
\newblock \emph{\bibinfo{journal}{Science}} \textbf{\bibinfo{volume}{354}},
  \bibinfo{pages}{1557--1562} (\bibinfo{year}{2016}).
\newblock \urlprefix\url{https://science.sciencemag.org/content/354/6319/1557}.

\bibitem{SRan2019Science}
\bibinfo{author}{Ran, S.} \emph{et~al.}
\newblock \bibinfo{title}{Nearly ferromagnetic spin-triplet superconductivity}.
\newblock \emph{\bibinfo{journal}{Science}} \textbf{\bibinfo{volume}{365}},
  \bibinfo{pages}{684--687} (\bibinfo{year}{2019}).
\newblock \urlprefix\url{https://science.sciencemag.org/content/365/6454/684}.

\bibitem{LJiao2020Nature}
\bibinfo{author}{Jiao, L.} \emph{et~al.}
\newblock \bibinfo{title}{{Chiral superconductivity in heavy-fermion metal
  UTe2}}.
\newblock \emph{\bibinfo{journal}{Nature}} \textbf{\bibinfo{volume}{579}},
  \bibinfo{pages}{523--527} (\bibinfo{year}{2020}).
\newblock \urlprefix\url{https://doi.org/10.1038/s41586-020-2122-2}.

\bibitem{Truncik2013NatComm}
\bibinfo{author}{Truncik, C. J.~S.} \emph{et~al.}
\newblock \bibinfo{title}{{Nodal quasiparticle dynamics in the heavy fermion
  superconductor CeCoIn5 revealed by precision microwave spectroscopy}}.
\newblock \emph{\bibinfo{journal}{Nat. Commun.}} \textbf{\bibinfo{volume}{4}},
  \bibinfo{pages}{2477} (\bibinfo{year}{2013}).
\newblock \urlprefix\url{https://doi.org/10.1038/ncomms3477}.

\bibitem{Klein1994PRB}
\bibinfo{author}{Klein, O.}, \bibinfo{author}{Nicol, E.~J.},
  \bibinfo{author}{Holczer, K.} \& \bibinfo{author}{Gr\"uner, G.}
\newblock \bibinfo{title}{{Conductivity coherence factors in the conventional
  superconductors Nb and Pb}}.
\newblock \emph{\bibinfo{journal}{Phys. Rev. B}} \textbf{\bibinfo{volume}{50}},
  \bibinfo{pages}{6307--6316} (\bibinfo{year}{1994}).
\newblock \urlprefix\url{https://link.aps.org/doi/10.1103/PhysRevB.50.6307}.

\bibitem{Hirschfeld1989PRB}
\bibinfo{author}{Hirschfeld, P.~J.}, \bibinfo{author}{W\"olfle, P.},
  \bibinfo{author}{Sauls, J.~A.}, \bibinfo{author}{Einzel, D.} \&
  \bibinfo{author}{Putikka, W.~O.}
\newblock \bibinfo{title}{Electromagnetic absorption in anisotropic
  superconductors}.
\newblock \emph{\bibinfo{journal}{Phys. Rev. B}} \textbf{\bibinfo{volume}{40}},
  \bibinfo{pages}{6695--6710} (\bibinfo{year}{1989}).
\newblock \urlprefix\url{https://link.aps.org/doi/10.1103/PhysRevB.40.6695}.

\bibitem{Hirschfeld1993PRL}
\bibinfo{author}{Hirschfeld, P.~J.}, \bibinfo{author}{Putikka, W.~O.} \&
  \bibinfo{author}{Scalapino, D.~J.}
\newblock \bibinfo{title}{Microwave conductivity of d-wave superconductors}.
\newblock \emph{\bibinfo{journal}{Phys. Rev. Lett.}}
  \textbf{\bibinfo{volume}{71}}, \bibinfo{pages}{3705--3708}
  (\bibinfo{year}{1993}).
\newblock \urlprefix\url{https://link.aps.org/doi/10.1103/PhysRevLett.71.3705}.

\bibitem{KZhang1994PRL}
\bibinfo{author}{Zhang, K.} \emph{et~al.}
\newblock \bibinfo{title}{{Measurement of the $\mathrm{ab}$ Plane Anisotropy of
  Microwave Surface Impedance of Untwinned
  ${\mathrm{YBa}}_{2}$${\mathrm{Cu}}_{3}$${\mathrm{O}}_{6.95}$ Single
  Crystals}}.
\newblock \emph{\bibinfo{journal}{Phys. Rev. Lett.}}
  \textbf{\bibinfo{volume}{73}}, \bibinfo{pages}{2484--2487}
  (\bibinfo{year}{1994}).
\newblock \urlprefix\url{https://link.aps.org/doi/10.1103/PhysRevLett.73.2484}.

\bibitem{Thiemann2018PRB}
\bibinfo{author}{Thiemann, M.}, \bibinfo{author}{Dressel, M.} \&
  \bibinfo{author}{Scheffler, M.}
\newblock \bibinfo{title}{{Complete electrodynamics of a BCS superconductor
  with $\ensuremath{\mu}\mathrm{eV}$ energy scales: Microwave spectroscopy on
  titanium at mK temperatures}}.
\newblock \emph{\bibinfo{journal}{Phys. Rev. B}} \textbf{\bibinfo{volume}{97}},
  \bibinfo{pages}{214516} (\bibinfo{year}{2018}).
\newblock \urlprefix\url{https://link.aps.org/doi/10.1103/PhysRevB.97.214516}.

\bibitem{Prozorov2006SST}
\bibinfo{author}{Prozorov, R.} \& \bibinfo{author}{Giannetta, R.~W.}
\newblock \bibinfo{title}{Magnetic penetration depth in unconventional
  superconductors}.
\newblock \emph{\bibinfo{journal}{Supercond. Sci. Technol.}}
  \textbf{\bibinfo{volume}{19}}, \bibinfo{pages}{R41} (\bibinfo{year}{2006}).
\newblock \urlprefix\url{http://stacks.iop.org/0953-2048/19/i=8/a=R01}.

\bibitem{Nakamine2019JPSJ}
\bibinfo{author}{Nakamine, G.} \emph{et~al.}
\newblock \bibinfo{title}{{Superconducting Properties of Heavy Fermion UTe2
  Revealed by 125Te-nuclear Magnetic Resonance}}.
\newblock \emph{\bibinfo{journal}{J. Phys. Soc. Jpn}}
  \textbf{\bibinfo{volume}{88}}, \bibinfo{pages}{113703}
  (\bibinfo{year}{2019}).
\newblock \urlprefix\url{https://doi.org/10.7566/JPSJ.88.113703}.

\bibitem{Gross1986ZPhyB}
\bibinfo{author}{Gross, F.} \emph{et~al.}
\newblock \bibinfo{title}{{Anomalous temperature dependence of the magnetic
  field penetration depth in superconducting UBe$_{13}$}}.
\newblock \emph{\bibinfo{journal}{Z. Phy. B}} \textbf{\bibinfo{volume}{64}},
  \bibinfo{pages}{175--188} (\bibinfo{year}{1986}).
\newblock \urlprefix\url{https://doi.org/10.1007/BF01303700}.

\bibitem{Ian2020arXiv}
\bibinfo{author}{Hayes, I.~M.} \emph{et~al.}
\newblock \bibinfo{title}{{Weyl Superconductivity in UTe$_2$}}.
\newblock \emph{\bibinfo{journal}{arXiv:2002.02539}}  (\bibinfo{year}{2020}).
\newblock \urlprefix\url{https://arxiv.org/abs/2002.02539}.

\bibitem{Radtke1993PRB}
\bibinfo{author}{Radtke, R.~J.}, \bibinfo{author}{Levin, K.},
  \bibinfo{author}{Sch\"uttler, H.-B.} \& \bibinfo{author}{Norman, M.~R.}
\newblock \bibinfo{title}{{Predictions for impurity-induced
  ${\mathit{T}}_{\mathit{c}}$ suppression in the high-temperature
  superconductors}}.
\newblock \emph{\bibinfo{journal}{Phys. Rev. B}} \textbf{\bibinfo{volume}{48}},
  \bibinfo{pages}{653--656} (\bibinfo{year}{1993}).

\bibitem{Mackenzie1998PRL}
\bibinfo{author}{Mackenzie, A.~P.} \emph{et~al.}
\newblock \bibinfo{title}{{Extremely Strong Dependence of Superconductivity on
  Disorder in ${\mathrm{Sr}}_{2}{\mathrm{RuO}}_{4}$}}.
\newblock \emph{\bibinfo{journal}{Phys. Rev. Lett.}}
  \textbf{\bibinfo{volume}{80}}, \bibinfo{pages}{161--164}
  (\bibinfo{year}{1998}).
\newblock \urlprefix\url{https://link.aps.org/doi/10.1103/PhysRevLett.80.161}.

\bibitem{Tristin2019PRB}
\bibinfo{author}{Metz, T.} \emph{et~al.}
\newblock \bibinfo{title}{{Point-node gap structure of the spin-triplet
  superconductor ${\mathrm{UTe}}_{2}$}}.
\newblock \emph{\bibinfo{journal}{Phys. Rev. B}}
  \textbf{\bibinfo{volume}{100}}, \bibinfo{pages}{220504}
  (\bibinfo{year}{2019}).
\newblock \urlprefix\url{https://link.aps.org/doi/10.1103/PhysRevB.100.220504}.

\bibitem{Huy2008Thesis}
\bibinfo{author}{Huy, N.~T.}
\newblock \emph{\bibinfo{title}{{Ferromagnetism, Superconductivity and Quantum
  Criticality in Uranium Intermetallics, Ph.D. thesis}}}
  (\bibinfo{publisher}{Universiteit van Amsterdam}, \bibinfo{year}{2008}).
\newblock
  \urlprefix\url{https://iop.fnwi.uva.nl/cmp/docs/huy/Thesis_NguyenThanhHuy_2008.pdf}.

\bibitem{Aoki2001Nat}
\bibinfo{author}{Aoki, D.} \emph{et~al.}
\newblock \bibinfo{title}{{Coexistence of superconductivity and ferromagnetism
  in URhGe}}.
\newblock \emph{\bibinfo{journal}{Nature}} \textbf{\bibinfo{volume}{413}},
  \bibinfo{pages}{613--616} (\bibinfo{year}{2001}).
\newblock \urlprefix\url{https://doi.org/10.1038/35098048}.

\bibitem{Sundar2019PRB}
\bibinfo{author}{Sundar, S.} \emph{et~al.}
\newblock \bibinfo{title}{{Coexistence of ferromagnetic fluctuations and
  superconductivity in the actinide superconductor ${\mathrm{UTe}}_{2}$}}.
\newblock \emph{\bibinfo{journal}{Phys. Rev. B}}
  \textbf{\bibinfo{volume}{100}}, \bibinfo{pages}{140502}
  (\bibinfo{year}{2019}).
\newblock \urlprefix\url{https://link.aps.org/doi/10.1103/PhysRevB.100.140502}.

\end{thebibliography}


\pagebreak
\clearpage
\newpage

\pagebreak
\clearpage
\newpage

\pagebreak
\clearpage
\newpage

\pagebreak
\clearpage
\newpage
\pagebreak


\section*{Supplementary material}

# Supplementary Information : Anomalous normal fluid response in a chiral superconductor UTe$_2$

**CONTENTS**

# SUPPLEMENTARY NOTE 1: MICROWAVE SURFACE IMPEDANCE MEASUREMENT PROCEDURE

## A. Transmission measurement procedure

The surface impedance $Z_s$ of a superconducting sample can be obtained from both resonant techniques and non-resonant techniques. In this work, $Z_s$ of the UTe$_2$ sample was measured by a resonant technique. Readers who are interested in the non-resonant techniques can refer to Ref. [1, 2]. For the resonant techniques, one can obtain $Z_s$ from the resonance properties (resonant frequency $f_0$ and quality factor $Q$) through either a microwave transmission or reflection measurement that involves the sample of interest. Here, we conducted a transmission measurement to obtain $Z_s$ of the sample.

For a microwave transmission measurement, the input and output microwave signals are generated and received by the Keysight vector network analyzer (VNA) N5242A (Supplementary Fig. 1). The VNA calculates a voltage ratio of the signal at its input (port 1) and output port (port 2) to obtain the complex microwave transmission scattering parameter $S_{21}^{\text{tot}}$ in the frequency domain. The generated signal from port 1 goes into the microwave transmission line (blue lines in Supplementary Fig. 1) of the BlueFors XLD dilution fridge and reaches down to the resonator attached to the mixing chamber plate (MXC) whose base temperature is 11 mK. The signal passed through the resonator comes back to port 2 of the VNA. The details of the two resonator designs used in this study will be discussed in Supplementary Note 2.

The $S_{21}^{\text{res}}$ measured by the VNA includes cable loss and phase shift from the transmission lines. For high-Q resonant techniques, the phase variation with frequency from the cable near the resonant frequency is typically small ($< 0.01$ radian for the 3dB bandwidth of the resonance in our experimental setup in Supplementary Note 2.D.), but the cable loss (decrease in the magnitude of $S_{21}$) can contribute a large portion to $|S_{21}^{\text{tot}}|$. One needs to subtract out the cable loss in order to obtain the magnitude of the scattering parameter

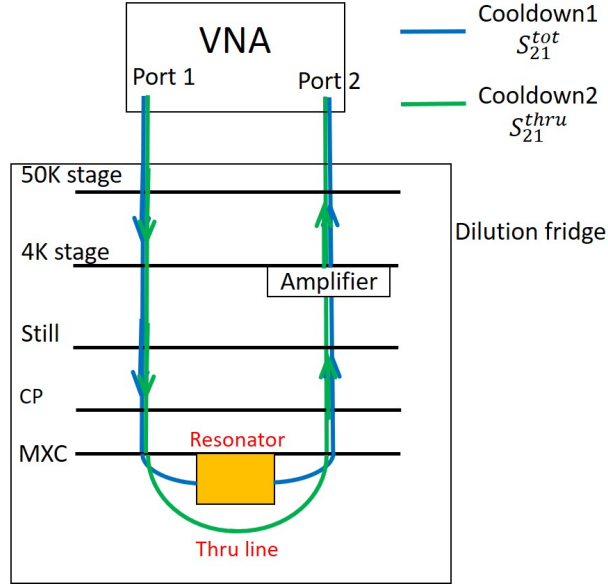

Supplementary Fig. 1. (Blue) schematic view of the signal path of the microwave transmission measurement $S_{21}^{\text{tot}}$ that includes the resonator. (Green) the signal path of the thru line transmission measurement $S_{21}^{\text{thru}}$ which shares the same signal path except that it bypasses the resonator, and is measured in a separate cooldown. The thru line measurement $S_{21}^{\text{thru}}$ gives a measure of the cable loss.

purely from the resonator. For this correction, an additional transmission line measurement called a thru line measurement is used. The thru line measurement follows the same signal path utilizing the same coaxial cables but bypasses the resonator, and is measured in a separate cooldown. The measured transmission from this correction line is called thru-$S_{21}$ ($S_{21}^{\text{thru}}$). By subtracting $|S_{21}^{\text{thru}}|$ from $|S_{21}^{\text{tot}}|$, one can estimate the magnitude of the resonator transmission $|S_{21}^{\text{res}}|$,

$$|S_{21}^{\text{res}}| = |S_{21}^{\text{tot}}| - |S_{21}^{\text{thru}}|. \tag{1}$$

With the above procedure for finding $S_{21}^{\text{res}}$, broadband transmission (6 ∼ 20 GHz, the range is limited by the microwave amplifier) is first measured to locate a resonant peak of interest. For the located peak, a narrow band measurement is done to precisely resolve the sharp $S_{21}^{\text{res}}(f)$ near the peak of the resonance.

### B. Determination of resonance properties $f_0$ and $Q$

With the measured resonator $S_{21}^{\text{res}}(f)$ curve, one can characterize the resonance properties. There is the resonant frequency $f_0$, which is related to the reactance of the resonator ($f_0 \sim 1/\sqrt{LC}$ where $L$ is the inductance and $C$ is the capacitance of a lumped-element equivalent circuit representation of the resonator), and the quality factor $Q$ which is related to the microwave dissipation $P_{\text{diss}}$ in the resonator ($1/Q \sim P_{\text{diss}}$). The extraction of $f_0$ and $Q$ can be done by fitting $S_{21}^{\text{res}}(f)$ curve near the resonance by a "Phase versus frequency fit" [3] which models the phase $\phi$ of the complex $S_{21}^{\text{res}}(f) = |S_{21}^{\text{res}}(f)|e^{i\phi(f)}$ with,

$$\phi(f) = \phi_0 + 2\tan^{-1}\left[2Q_L\left(1 - \frac{f}{f_0}\right)\right], \tag{2}$$

where $Q_L$ is a "loaded" quality factor and $\phi_0$ is a phase offset.

Ideally, the inverse of the quality factor $1/Q$ should describe the dissipation from the resonator itself. However, to excite and pick up the microwave signal, the resonator should be coupled to the external world through excitation/pick-up loops. This coupling causes some portion of the energy in the resonator to leak out to the external world [4], adding an additional source of energy dissipation which is described by $1/Q_c$ where $Q_c$ is called the coupling $Q$. The $Q$ arising purely from losses within the resonator is called the unloaded $Q$ ($Q_u$). The $Q$ factor obtained from fitting the raw $S_{21}$ data above is called the loaded $Q$ ($Q_L$) because it describes the dissipation from both the resonator itself $1/Q_u$ and the coupling $1/Q_c$,

$$1/Q_L = 1/Q_u + 1/Q_c. \tag{3}$$

To obtain information on the microwave dissipation purely from the resonator, one needs to know $Q_u$ which removes the effect of coupling. In the case of equal coupling from the input and output side, $Q_u$ can be calculated as [5, 6],

$$Q_u = \frac{Q_L}{1 - |S_{21}(f_0)|}. \tag{4}$$

To obtain the dissipated microwave power from the sample $P_{\text{diss}}^s$, which gives the resistance $R_s$ of the sample, one needs to further subtract the dissipated power from the background components $P_{\text{diss}}^b$ of the resonator. Note that the typical sources of the background dissipation of the resonator are Ohmic dissipation from the enclosure of the resonator (if it is normal metal), dielectric loss from the dielectric materials, and radiation loss from any

openings of the resonator. The resonator dissipation and the unloaded Q can be written as,

$$P_{\text{diss}}^{\text{tot}} = P_{\text{diss}}^{s} + P_{\text{diss}}^{b}, \tag{5}$$

$$\frac{1}{Q_u} = \frac{1}{Q_s} + \frac{1}{Q_b}. \tag{6}$$

To remove the background contribution on the dissipation and Q, two consecutive measurements are required. One is the measurement without the sample ($Q_u^{\text{w.o.}}$) which measures the background dissipation. The second one is the measurement with the sample inserted in the resonator ($Q_u^{\text{w.}}$) which now includes the actual sample properties. The sample quality factor $Q_s(T)$ can be finally determined by the results of those two measurements [7],

$$1/Q_u^{\text{w.o.}} = 1/Q_b, \tag{7}$$

$$1/Q_u^{\text{w.}} = 1/Q_b + 1/Q_s, \tag{8}$$

$$1/Q_s(T) = 1/Q_u^{\text{w.}}(T) - 1/Q_u^{\text{w.o.}}(T). \tag{9}$$

Once the subtraction is done, it leaves only the sample contribution to the quality factor $Q_s$.

## C. Converting resonance properties ($f_0$ and $Q$) into surface impedance ($Z_s$)

Once $f_0(T)$ and $Q_s(T)$ are obtained, the surface impedance $Z_s = R_s + iX_s$ of the sample can be determined as [8],

$$R_s(T) = G_{\text{geo}}/Q_s(T) \tag{10}$$

$$\Delta X_s(T) = -2G_{\text{geo}} \frac{\Delta f_0(T)}{f_0(T_{\text{min}})}. \tag{11}$$

$$G_{\text{geo}} = \frac{\omega \mu_0 \int_V dV |H(x,y,z)|^2}{\int_S dS |H(x,y)|^2}. \tag{12}$$

Here, $\Delta f_0(T) = f_0(T) - f_0(T_{\text{min}})$, $\Delta X_s(T) = X_s(T) - X_s(T_{\text{min}})$ where $T_{\text{min}}$ is the minimum temperature of the measurement, $\int_V dV$ represents the integration over the volume of the resonator, and $\int_S dS$ represents the integration over the surface of the sample. $G_{\text{geo}}$ is the sample geometric factor which can be calculated either analytically using the field solution inside the resonator for each resonant mode (if the solution is available), or experimentally by comparing the experimentally measured $R_s$ from Supplementary Equation (10) with the calculated $R_s = \sqrt{\mu \omega \rho_{\text{dc}}/2}$ from electrical resistivity $\rho_{\text{dc}}$ of a dc transport measurement in

the normal state. Note that this equality is valid in the large scattering rate limit ($\omega\tau \ll 1$. Here, $\omega = 2\pi f$ and $\tau$ is the scattering lifetime of the electron) in the normal state of the sample [9]. This limit can be reached by increasing the sample temperature well above $T_c$ which typically reduces the scattering lifetime $\tau$. Once the limit is reached, $G_{\text{geo}}$ can be determined.

Also, the absolute value of $X_s(T)$ can be determined in the following way. When the $\omega\tau \ll 1$ regime is reached in the normal state, the complex conductivity $\tilde{\sigma} = ne^2\tau/m^*(1 + i\omega\tau) \approx ne^2\tau/m^*$ only has a real part. As $Z_s = \sqrt{i\mu\omega/\tilde{\sigma}}$, purely real $\tilde{\sigma}$ results in $R_s = X_s = \Delta X_s(T) + X_{s0}$. From the comparison of $R_s$ and $\Delta X_s$, the offset in the sample reactance $X_{s0}$ can be found, which yields the absolute value of $X_s(T)$ for the full temperature range. With the absolute values of $R_s(T)$ and $X_s(T)$ determined, one can calculate the microwave complex conductivity $\tilde{\sigma}(T)$ whose real and imaginary parts provide useful information about the normal fluid and superfluid response of a superconductor.

One important question here is, what is the unambiguous experimental signature that one has achieved the $\omega\tau \ll 1$ regime in the normal state? Suppose the $\omega\tau \ll 1$ regime is achieved above a certain temperature $T_{\text{ls}} > T_c$. For $T > T_{\text{ls}}$, $R_s(T) = X_s(T)$ so their temperature derivative should also satisfy $dR_s/dT = dX_s/dT$ [10]. This means that once a temperature above which $dR_s/dT = d(\Delta X_s)/dT$ is satisfied (note that $dX_s/dT = d(\Delta X_s)/dT$), $T_{\text{ls}}$ can be identified. Note that the relative magnitude of the temperature derivative of $R_s$ and $X_s$ has $G_{\text{geo}}$ as a common factor, thus the value of $G_{\text{geo}}$ does not affect this comparison test.

## SUPPLEMENTARY NOTE 2: TYPES OF MICROWAVE RESONATORS

### D. Disk dielectric resonator

To measure the surface impedance $Z_s$ and the corresponding microwave complex conductivity $\tilde{\sigma}$ of the UTe$_2$ single crystal sample with the procedures described in Supplementary Note 1, two types of resonator design were used. The first type is the disk dielectric resonator (Supplementary Fig. 2). The disk dielectric resonator (disk DR) consists of a cylindrical rutile disk with a diameter of 3 mm, and height of 2 mm. With a high in-plane dielectric constant over 110 below 20 K [11], the rutile disk facilitates its fundamental resonant mode (TE$_{011}$) at $f_0 \approx 11$ GHz [12]. In this mode, the disk induces a radial microwave magnetic

field on the sample surface, which induces an azimuthal circulating current on the sample surface. Here, the UTe$_2$ sample is mounted with an *ab*-plane facet in contact with the top of the rutile disk. Therefore, the measured $Z_s$ data in the main text represents *ab*-plane electrodynamics. Note that to avoid the effect of microwave heating, we measured $f_0$ as a function of input microwave power $P_{in}$, and found that $f_0$ is independent of $P_{in}$ below $-30$ dBm (measured at the source). We used $P_{in} = -40$ dBm to produce the $Z_s(T)$ data. In this case, the injected microwave power at the input coupling loop of the resonator was $-51$ dBm. Electromagnetic simulation (HFSS) of the resonator and sample provides an estimation of the microwave magnetic field intensity at the sample surface of $\approx 1\mu$T.

To cut down the background dissipation and obtain a resonant microwave transmission with a higher qualify factor, the top and the bottom side of the disk is enclosed by a niobium plate and film. This niobium top and bottom plate structure significantly reduces the background Ohmic loss because they are superconducting, and they have temperature-independent properties in the superconducting temperature range for UTe$_2$. Also the semi-cavity niobium structure reduces the radiation loss to a marginal level. The cutoff frequency of wave propagation through the $\approx 3$ mm diameter aperture and the side space is 58.6 GHz. These cutoffs are much higher than the measurement frequency of 11 GHz, and hence the radiation field cannot propagate. Lastly, the dielectric loss should be marginal too since the loss tangent $\tan\delta = 1/Q_{\text{diel}}$ is $\sim 10^{-8}$ for the rutile below 10 K [11]. With the design to reduce the background loss mechanism, the high background quality factor $Q_b$ (low background loss) was obtained $Q_b \approx 10^5$. This $Q_b$ was subtracted from the measured $Q_u$ with the sample present (Supplementary Fig. 2(b)), as described in Supplementary Equation 9.

Once the background measurement was done, the sample was inserted and thermally glued by N-grease. The temperature dependent resonance properties $\Delta f_0(T) = f_0(T) - f_0(50\text{mK})$ and $Q(T)$ which involves the sample were measured for the temperature range of 50 mK to 3 K. The temperature range is limited because above 3 K, the niobium semi-cavity structure starts to show finite temperature dependence on its contribution to $\Delta f_0$ and $Q$.

Once the $\Delta f_0(T)$ and $Q_s(T)$ were determined, they were converted to $R_s(T)$ and $\Delta X_s(T)$ with temporary value of $G_{geo} = 1$. To determine the absolue value of the $X_s(T)$, one needs to find the temperature range of the $\omega\tau \ll 1$ regime in the normal state. To do so, $R_s(T)$ and $\Delta X_s(T)$ curves were compared and overlaid. As one can see from Fig. 1 of the main text,

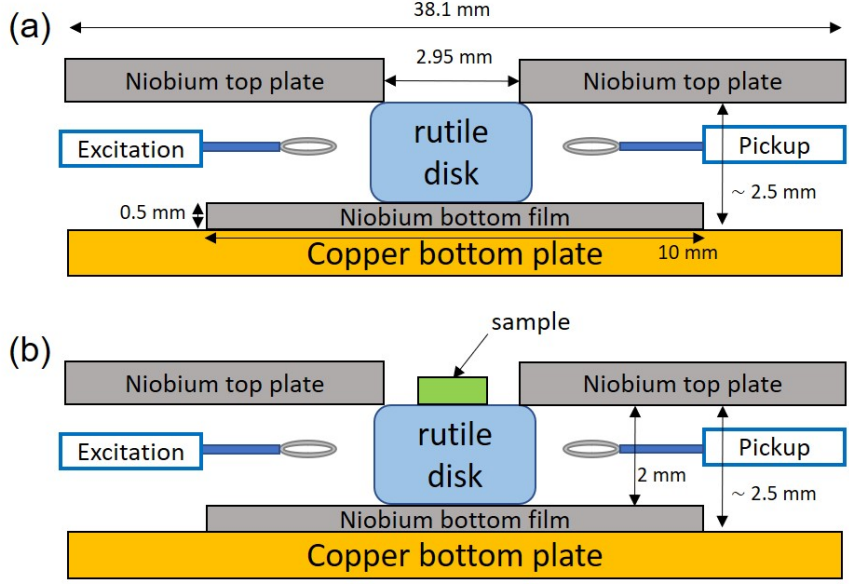

Supplementary Fig. 2. Cross-section view of the cylindrically-symmetric disk dielectric resonator setup to facilitate $TE_{011}$ resonance mode at $\approx 11$ GHz (a) without the sample for the background measurement (b) with the sample for the sample $Z_s$ measurement.

both curves show the very same slope in the 2 to 3 K range. This suggests that the $\omega\tau \ll 1$ regime is reached, and the absolute value of $X_s(T)$ was determined by $R_s(T) = X_s(T)$ between 2 and 3 K. Then, the $R_s$ at 3 K, which involves the undetermined $G_{geo}$, was compared to the $R_s$ calculated from the electrical resistivity obtained from the dc transport measurement [13]. Note that there is 1:2 anisotropy in the value of the resistivity along the $a$ and $b$-axis in the above temperature range. The two resistivity values were averaged with the weighting factor ratio 3:7, which is an inverse of the ratio of the dimensions of the sample in each direction. In this way, $G_{geo}$ was determined to be 713 $\Omega$.

However, one may be concerned that the 2 to 3 K interval is not wide enough to correctly identify the convergence between the slope of the $R_s(T)$ and $X_s(T)$ curves. For a more rigorous check, we adopted the hollow cylinder dielectric resonator design similar to Ref. [7] with the hot finger technique [14–16] to measure $R_s(T)$ and $X_s(T)$ up to 20 K, where the $\omega\tau \ll 1$ regime is safely reached in the cases of several other heavy fermion superconductors [17, 18].

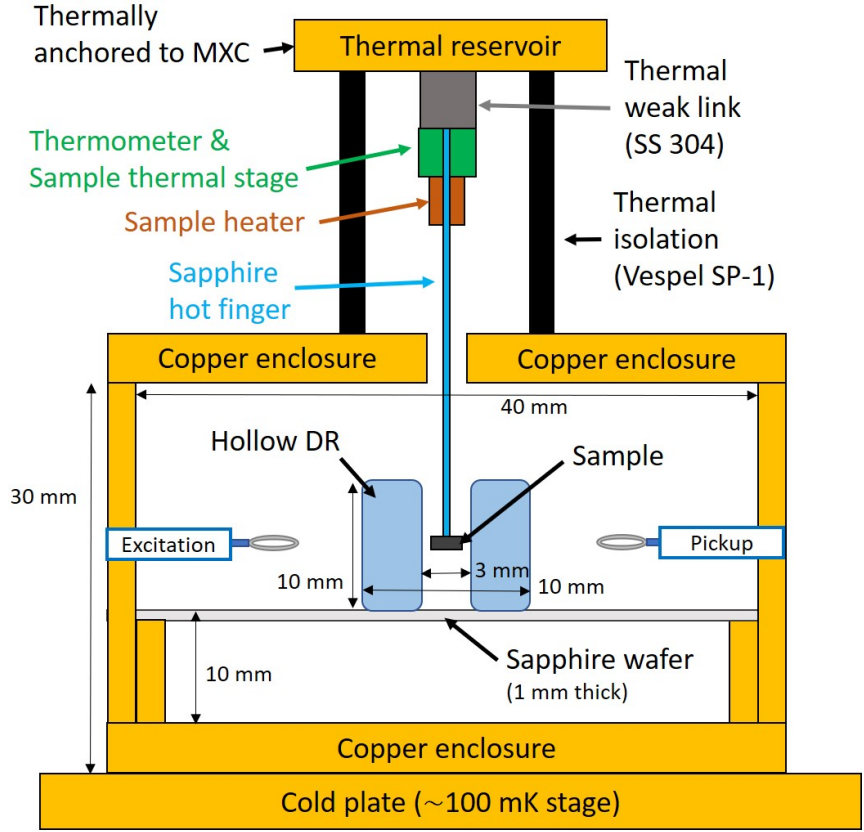

Supplementary Fig. 3. Cross-sectional view of hollow rutile cylinder dielectric resonator (HDR) and the attached sapphire hot finger setup. The base temperature of the sample thermal stage is ∼220 mK.

### E. Hollow cylinder dielectric resonator (HDR) with hot finger

The key for measuring $Z_s(T)$ of the sample in the normal state up to a high temperature where $\tau$ becomes small enough (i.e. 20 K) is to fix the background contribution to the $f_0$ and $Q$ while the sample temperature is varied. To accomplish this goal, two main features are introduced in the HDR. First, the niobium structure, which contributes a strong temperature dependence to $f_0$ and $Q$ above 3 K, was replaced with copper walls (Supplementary Fig. 3). Copper shows only a small change ($< 3$ %) in its electrical resistivity between 4 and 20 K [19], cutting down the background temperature dependence on $Q$ to a manageable level. The copper walls were moved away from the sample and the rutile to minimize the contribution from the walls to the quality factor.

Secondly, the hot finger technique [14–16] was introduced to decouple the sample stage

9

temperature and the resonator background temperature (rutile, copper walls). This temperature decoupling was designed especially because the rutile cylinder has a finite temperature dependence in its dielectric constant above 10 K [11], which can give a background temperature dependence in $f_0$. The hot finger structure utilizes Vespel SP-1 rods to thermally isolate the resonator (which is directly thermally anchored to the cold plate at $\sim$ 100 mK) and the thermal reservoir (anchored to the mixing chamber of the fridge with thermal straps). The thermal reservoir and the sample thermal stage are weakly connected by a non-magnetic stainless steel tube (304 SS). This "weak-link" provides a certain degree of thermal decoupling during the sample temperature ramping, while the cooldown time is still in a reasonable range. A 1 mm diameter single crystal sapphire rod is then attached to the sample thermal stage. At the other end of the sapphire rod, the sample is mounted with GE varnish, and placed in the middle of the hollow rutile cylinder where the microwave magnetic field intensity is the highest. The 3 mm diameter cylindrical bore of the 10 mm outer diameter and 10 mm height hollow rutile cylinder ensures no mechanical, and hence no thermal contact between the sample and cylinder. The thermometer is attached to the sample thermal stage and the resistive heater is wound around the sapphire rod to increase the sample temperature. The heater wires are twisted to cancel the dc magnetic field from the current through them. With this hot finger design, when the sample temperature is increased to 20 K, the resonator temperature only increases to 620 mK. With the two changes made in the HDR design, it successfully decouples and minimizes the background contribution to the resonance properties. The data taken from the HDR is discussed in Supplementary Note 3.

**SUPPLEMENTARY NOTE 3: NORMAL STATE $Z_s(T)$ AND SCATTERING LIFETIME**

With the UTe$_2$ sample (the same one used in the disk DR) mounted in the HDR, the $Z_s$ data of the sample was obtained from 220 mK to 20 K. The resonant mode under study (TE$_{025}$) has a resonant frequency $f_0$ at $\approx$ 9.07 GHz, which is the closest to $f_0$ of the resonant mode of the disk DR (11 GHz). The obtained $Z_s$ is displayed in Supplementary Fig. 4. As one can see, the slope of the $R_s$ and $X_s$ curves are already quite similar above $T_c$, and completely merge above 5 K conservatively speaking, and do not diverge from each other up to 20 K, implying the $\omega\tau \ll 1$ regime is safely reached.

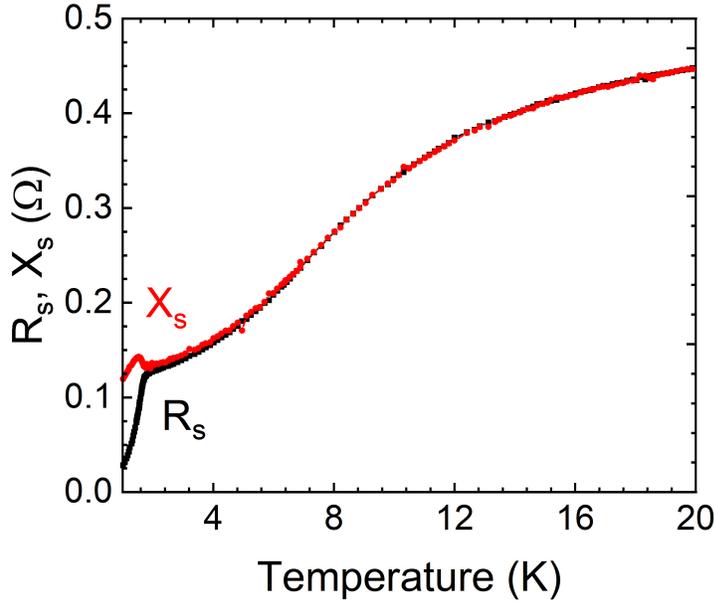

Supplementary Fig. 4. Measured $Z_s(T)$ data at 9.07 GHz of the UTe$_2$ sample obtained from the HDR.

In the normal state, from the simple Drude model one can estimate the scattering lifetime $\tau$ from $Z_s$,

$$\frac{i\mu\omega}{Z_s^2} = \tilde{\sigma} = \frac{ne^2}{m^*}\frac{\tau}{1+i\omega\tau}. \tag{13}$$

Here, $\mu$ is the permeability, $n$ is the electron density, $m^*$ is the effective mass of the electron. One can regard the model as having two unknowns ($ne^2/m^*$, $\tau$) and two equations (real and imaginary part), thus it is solvable at each temperature. The oscillator strength $ne^2/m^* = 2.33\times 10^{18}$ C$^2$m$^{-3}$kg$^{-1}$ is obtained. The obtained $\omega\tau(T)$ above $T_c$ is shown in Supplementary Fig. 5. In the normal state above $T_c$, $\omega\tau \leq 0.05$ ($\tau \leq 0.88$ ps), satisfying the large scattering limit. Therefore, equating $R_s = X_s$ above 2 K ($\approx 1.25T_c$) in the case of the disk DR in Supplementary Note 2.D is further justified to a good approximation.

Note that the large scattering rate in the normal state $\Gamma_n(T > T_c) = 1/\tau(\sim 1$ THz$) \gg 2\Delta_0(0)(\sim 130$ GHz$)$ estimated above does not contradict the small bulk impurity scattering rate at low temperature $\Gamma_{\text{imp}}(0) \leq 0.06\Delta_0(0)$ obtained from the superfluid density analysis (Discussion in the main text). The scattering rate in a superconductor consists of elastic scattering such as impurity scattering $\Gamma_{\text{imp}}$, and inelastic scattering such as electron-electron scattering $\Gamma_{\text{ee}}$ (e.g. due to spin fluctuations), electron-phonon scattering $\Gamma_{\text{e-ph}}$ and other

11

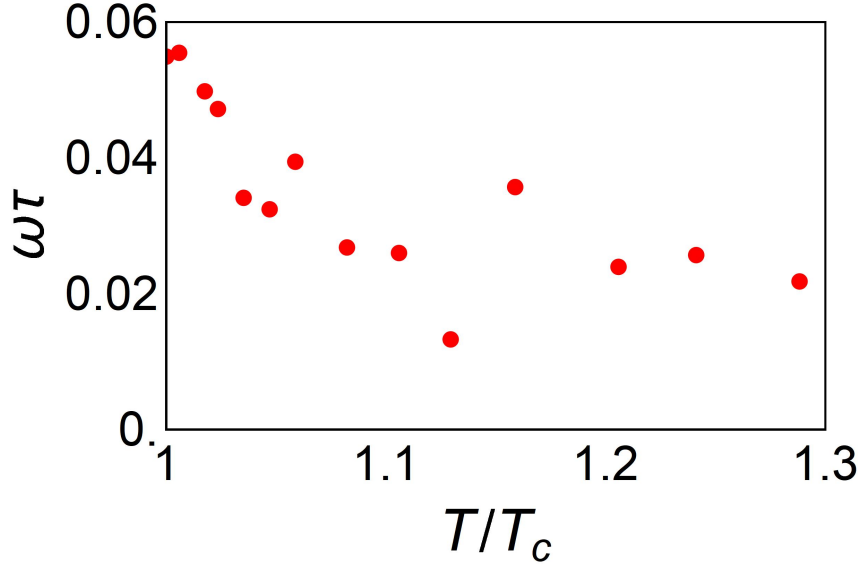

Supplementary Fig. 5. Extracted scattering time $\omega\tau$ from a Drude model analysis to the normal state complex conductivity data.

possible mechanisms. The experimentally obtained bulk scattering rate $\Gamma_{\text{tot}}$ should be the sum of these various scattering rates $\Gamma_{\text{tot}} = \Gamma_{\text{imp}} + \Gamma_{\text{ee}} + \Gamma_{\text{e-ph}} + \Gamma_{\text{others}}$.

An important point here is that the scattering rates from different mechanisms can follow different temperature dependences across $T_c$. For example, while $\Gamma_{\text{imp}}$ does not show noticeable difference across the $T_c$ as seen from Ti (430/490 GHz for below/above $T_c$ [20]), $\Gamma_{\text{ee}}$ from spin fluctuations shows a very sharp drop just below $T_c$ with a $T^3$ dependence. This sharp drop in $\Gamma_{\text{ee}}$ was theoretically predicted [21] and experimentally observed in spin fluctuation dominating superconductors such as YBCO [22] and CeCoIn$_5$ [23]. As a result, for systems where the inelastic scattering mechanism dominates $\Gamma_{\text{ee}} \gg \Gamma_{\text{imp}}$ in the normal state, the ratio of the scattering rates in the normal and superconducting state ($\Gamma(T_c)/\Gamma(0)$) can reach up to 40 (CeCoIn$_5$) and 400 (YBCO). In this case, it is compatible to have a large total scattering rate in the normal state $\Gamma_{\text{tot}}(T_c) = \Gamma_{\text{imp}} + \Gamma_{\text{ee}}(T_c) \approx \Gamma_{\text{ee}}(T_c) \gg 2\Delta_0(0)$, and to have a low impurity scattering rate in the superconducting state $\Gamma_{\text{tot}}^{\text{bulk}}(0) \approx \Gamma_{\text{imp}} \ll 2\Delta_0(0)$ at the same time.

UTe$_2$ is a system with the presence of strong magnetic fluctuations in the normal state as shown from the muon spin relaxation study and critical scaling behavior of the magnetism [13, 24]. Therefore, in UTe$_2$, it is possible that spin fluctuations dominate the scattering mechanism in the normal state $\Gamma_{\text{ee}}(T_c) \gg \Gamma_{\text{imp}}$. In this case, as shown in the above para-

graph, having a large $\Gamma_{\text{tot}}(T_c) \gg 2\Delta_0(0)$ in the normal state and being in the clean limit in the superconducting state $\Gamma_{\text{imp}} \ll 2\Delta_0(0)$ can be reconciled due to the sharp drop in $\Gamma_{\text{ee}}$ just below $T_c$.

## SUPPLEMENTARY NOTE 4: ESTIMATION FOR THE MEAN FREE PATH AND JUSTIFICATION FOR LOCAL ELECTRODYNAMICS

When the measured surface impedance $Z_s$ is converted to the complex conductivity, the simple relation $Z_s = \sqrt{i\mu_0\omega/\tilde{\sigma}}$ is used. This relation is valid in the local electrodynamics regime, which is $\xi(0) \ll \lambda(0)$ in the superconducting state and $l_{\text{mfp}} \ll \delta_{\text{skin}}$ in the normal state. Here, $\xi(0)$ is the superconducting coherence length at $T = 0$, $\lambda(0)$ is the magnetic penetration depth at $T = 0$, $l_{\text{mfp}}$ is the mean free path, and $\delta_{skin}$ is the skin depth. To see if UTe$_2$ is indeed in this local regime, one needs to compare the values of $l_{\text{mfp}}$, $\delta_{\text{skin}}$, $\xi(0)$, and $\lambda(0)$. First, for the superconducting state, $\xi(0)$ and $\lambda(0)$ are compared. Considering the estimated $\xi(0) = 4 \sim 7$ nm from $H_{c2}$ measurement [13, 25] and $\lambda(0) \approx 791$ nm from the main text, we see that $\xi(0) \ll \lambda(0)$.

Second, for the normal state, since the scattering lifetime $\tau_n(T \geq T_c) \leq 0.88$ ps is obtained from Supplementary Fig. 5 (Supplementary Note 3), once the Fermi velocity $v_F$ is known, $l_{\text{mfp}} = v_F \tau_n$ can be easily estimated. In the recent angle-resolved photoemission spectroscopy (ARPES) study [26], the band Fermi velocity $v_F^{\text{band}} \approx 3.9 \times 10^5$ m/s was obtained for the band along the $k_x$ direction, which dominates the ARPES signal intensity. $l_{\text{mfp}}$ estimated from this Fermi velocity is $\approx 340$ nm. The normal state skin depth can be estimated from the electrical resistivity $\rho_{\text{tr}}$ ($\delta_{\text{skin}} = \sqrt{2\rho_{\text{tr}}/\omega\mu}$). Using the ab-axis average of the $\rho_{\text{tr}}$ reported in Ref. [13], $\delta_{\text{skin}}$ is estimated to be 2.74 $\mu$m at 1.8 K and 9.07 GHz, and hence $l_{\text{mfp}} \ll \delta_{\text{skin}}$. Therefore, UTe$_2$ is in the local electrodynamics regime both in the superconducting and normal state.

13

# SUPPLEMENTARY NOTE 5: INCONSISTENCY BETWEEN SINGLET PAIRING STATES AND LOW TEMPERATURE BEHAVIOR OF THE PENETRATION DEPTH

As discussed in the Methods section, the penetration depth of UTe$_2$ shows a low temperature behavior which can be fit to $\Delta\lambda_{\text{eff}}(T) = aT^c$ with $c = 2.11$. Here, the temperature exponent $c$ is a good indicator for the nodal structure of the superconducting gap, which is closely related to the pairing state. In this section, the compatibility between the penetration depth data $\Delta\lambda_{\text{eff}}(T)$ and various singlet pairing states will be examined.

The representative singlet pairing states are fully gapped $s$ and line-nodal $d$-wave pairing states. First, the fully gapped $s$-wave state is expected to show an exponential temperature dependence with $c \geq 4$ power law at low temperature $T/T_c < 0.3$ [27] Second, a line-nodal $d_{x^2-y^2}$-wave pairing state is expected to show linear ($c \sim 1$) temperature dependence in the clean, local electrodynamics limit. Both of which are inconsistent with our data.

However, for the case of $d_{x^2-y^2}$-wave state, there are the possible cases which show a stronger than a linear temperature dependence (so that it is close to the observed quadratic behavior). The possible cases are a $d_{x^2-y^2}$ pairing state in the dirty limit [28], and $d_{x^2-y^2}$ state in the nonlocal electrodynamics regime [29]. In both cases, the low temperature behavior of the penetration depth shows a quadratic behavior below a crossover temperature $T^*$, and shows a linear behavior above $T^*$. This crossover behavior can be expressed as [28, 29],

$$\Delta\lambda_{\text{eff}}(T) = \alpha \frac{T^2}{T + T^*}, \qquad (14)$$

and hence the value of the temperature exponent for $\Delta\lambda_{\text{eff}}(T)$ ranges from $1 \sim 2$ depending on the value of $T^*$.

The possibility of the dirty $d_{x^2-y^2}$-wave scenario can be examined by checking the value of $T^*$. To obtain a temperature dependence closed to quadratic, $T^*_{\text{imp}}$ has to be much larger than $T_c$. For example, from Supplementary Equation 14, $T^*_{\text{imp}} \geq 4T_c$ is required to obtain a power law exponent of $c \geq 1.95$. $T^*_{\text{imp}}$ can be converted to an equivalent impurity scattering rate $\Gamma_{\text{imp}} \simeq (k_B T^*_{\text{imp}})^2/(0.83^2 \Delta_0(0))$ [27]. If one assumes $\Delta_0(0) = 2.14 k_B T_c$ which is a weakly-coupled BCS result of $d_{x^2-y^2}$ state, $T^*_{\text{imp}} \geq 4T_c$ means $\Gamma_{\text{imp}} \geq 10.86 k_B T_c$. However a bulk impurity scattering rate much larger than the critical temperature will suppress the superconducting state even at zero temperature for a nodal superconductor [30, 31].

Therefore, the dirty limit $d_{x^2-y^2}$ pairing scenario is inconsistent with our data.

The possibility of a nonlocal electrodynamic regime can also be examined by the value of $T^*$ which can be estimated from a superconducting coherence length $\xi(0)$, penetration depth $\lambda_{\text{eff}}(0)$, and the gap energy $\Delta(0)$ at zero temperature by $T^*_{\text{nonlocal}} = \Delta(0)\xi(0)/\lambda_{\text{eff}}(0)$ [29]. This estimation gives $T^*_{\text{nonlocal}} \approx 0.01 T_c$ for our sample. This means that the penetration depth should follow the linear temperature dependence above $0.01 T_c$, which is inconsistent with our data, which shows a quadratic behavior up to the full fitting range $0 \sim 0.3 T_c$.

Another possibility for the $d$-wave pairing state other than $d_{x^2-y^2}$ is the $d_{zx}+id_{yz}$ pairing state. This pairing state possesses both point nodes and a line node. For example, one can imagine point nodes along the north and south pole, and a line node at the equatorial line of the superconducting Fermi surface. In this case, if the current in the sample is induced only along the point nodal direction, it may show the quadratic temperature dependence even in the presence of the line node. However, as discussed in Supplementary Note 2.D, the disk DR induces a circulating current in all directions in the $ab$-plane of the sample surface. Therefore, the response from the line nodal direction, which gives a linear $\Delta\lambda_{eff}(T)$, cannot be missed.

As just discussed. the above possibilities of representative singlet pairing states turn out to be inconsistent with our data. Also, the singlet pairing state cannot explain the reported upper critical field $H_{c2}$ which is larger than the paramagnetic limiting field [13], and cannot explain the absence of a change in the NMR (nuclear magnetic resonance) Knight shift across $T_c$. [13, 32]. Therefore, in the superfluid density analysis of the main text, only spin-triplet pairing states are considered.

**SUPPLEMENTARY NOTE 6: FURTHER QUALITATIVE EVIDENCE OF THE RESIDUAL NORMAL FLUID THROUGH THE TWO-FLUID MODEL**

To further understand the anomalous residual $\sigma_1$, a two-fluid model for the electrodynamics can be introduced. For a homogenous superconducting medium, the two-fluid model describes the complex conductivity of the superconductor in terms of the superfluid and normal fluid contributions [17, 33],

$$\tilde{\sigma}(T) = \frac{ne^2}{m^*}\left(\frac{f_s(T)}{i\omega} + \frac{f_n(T)\tau_{\text{tot}}(T)}{1+i\omega\tau_{\text{tot}}(T)}\right), \quad (15)$$

where $n$ is total electron density, $m^*$ is the effective mass of the electrons, $f_n$ is normal fluid fraction, $f_s = 1 - f_n$ is the superfluid fraction, and $\tau_{\text{tot}}$ is the effective normal fluid scattering lifetime. Note that $1/\tau_{\text{tot}}$ sums up scattering rates of various origins in a weighted manner, such as bulk impurity $\Gamma_{imp}^{bulk}$ and topological surface normal fluid $\Gamma^{surf}$ (if there is any).

From Supplementary Equation (15), the real and imaginary parts of the conductivity can be obtained as,

$$\sigma_1 = \frac{ne^2}{m^*} \frac{f_n \tau_{\text{tot}}}{1 + \omega^2 \tau_{\text{tot}}^2} \tag{16}$$

$$\sigma_2 = \frac{ne^2}{m^*} \frac{1 + \omega^2 \tau_{\text{tot}}^2 - f_n}{\omega(1 + \omega^2 \tau_{\text{tot}}^2)}. \tag{17}$$

If one divides Supplementary Equation (17) by (16), it yields a quadratic equation for $\omega \tau_{\text{tot}}$ in terms of the conductivity ratio $\sigma_2/\sigma_1$ and normal fluid fraction $f_n$ with solution,

$$\omega \tau_{\text{tot}} = 0.5 \left[ \frac{\sigma_2}{\sigma_1} f_n \pm \sqrt{\left(\frac{\sigma_2}{\sigma_1} f_n\right)^2 - 4(1 - f_n)} \right]. \tag{18}$$

To have real solutions for $\tau_{\text{tot}}$, $f_n$ must satisfy the following condition,

$$f_n \geq \frac{2}{1 + \sqrt{1 + (\sigma_2/\sigma_1)^2}} \tag{19}$$

If UTe$_2$ is a conventional superconductor without residual normal fluid, $\sigma_2/\sigma_1 \to \infty$ as $T \to 0$ [23, 34, 35], which yields $f_n \geq 0$, allowing the normal fluid to be fully depleted. However this is not the case for UTe$_2$. The complex conductivity data from the UTe$_2$ sample shows a finite value of $\sigma_2(0)/\sigma_1(0) = 2.22$. This value sets a non-vanishing lower limit for the residual normal fluid fraction $f_n(0) \geq 58$ %, not allowing full depletion of the normal fluid in the $T \to 0$ limit. Therefore, a conventional superconductor scenario cannot explain the $\tilde{\sigma}$ of UTe$_2$.

One important warning is that the above estimation of the residual normal fluid fraction should not be taken quantitatively, but instead, should be taken as qualitative evidence of the residual normal fluid. The reason is that the two-fluid model applies for a homogeneous medium, and requires modification for an inhomogeneous medium. For example, if one attributes the residual $\sigma_1/\sigma_n$ to the surface normal fluid of a chiral superconductor and wants a more quantitatively accurate analysis, one needs to introduce a three-fluid model [18].

$$\tilde{\sigma}(\omega, T) = \frac{ne^2}{m^*} \left( \frac{f_s(T)}{i\omega} + \frac{f_{nb}(T)\tau_b(T)}{1 + i\omega\tau_b(T)} + \frac{f_{ns}(T)\tau_s(T)}{1 + i\omega\tau_s(T)} \right), \tag{20}$$

16

where $b$ stands for bulk, and $s$ stands for surface. However, this is an underdetermined system of equations (2 equations with 4 unknowns). Thus, one needs continuous frequency dependence data of the conductivity $\tilde{\sigma}(\omega)$, so that those unknowns can be estimated from fitting the frequency dependence. A microwave Corbino measurement on thin film samples [1, 2] is encouraged in the future for this purpose.

Also, the spatial profile of the microwave magnetic field and current inside the single crystal sample is complicated to quantitatively estimate. For example, if there are edges, the enhanced current distribution along the edges [36, 37] should be also considered to obtain a quantitatively rigorous estimation. For the case of a homogeneous superconductor, the spatial profile does not affect the procedure of estimating $f_n$ and $\tau$ since these quantities are assumed to be uniform across the medium. However, for an inhomogeneous case such as the surface normal fluid and bulk superfluid, the complicated spatial profile of the field is a problem since it determines the weighting factor of the surface and bulk response. A simplified spatial profile, along with systematic control on it, and hence more quantitative study of the properties of the surface normal fluid, can be accomplished with development of superconducting thin films of $UTe_2$.

## SUPPLEMENTARY NOTE 7: REMARKS ON THE RESIDUAL SPECIFIC HEAT RESULT

In the discussion of the main text, the absence of the linear term in the bulk thermal conductivity ($\kappa/T$) is cited as evidence of the absence of the bulk normal fluid [38]. However, it may be confusing to the readers because the initial specific heat study in Ref. [13] showed a residual linear term $c/T$ which was, at that time, presented as a signature of bulk normal fluid. Here, we provide clarification on the result and interpretation of the $c/T$ study.

The initial $c/T$ data was taken above 400 mK. The later study [38] went down below 100 mK, and revealed that the high residual $c/T$ value is due to the existence of a $1/T^{0.33}$ divergent term in $c/T$ as $T \to 0$. The later study also revealed that the temperature dependence of the raw specific heat data ($c/T$ vs $T$) does not obey the entropy balance condition which a superconductor must satisfy with this divergent term. Once the contribution from the divergent term in $c/T$ is subtracted, the entropy balance is perfectly recovered. This observation implies that the divergent term in $c/T$ as $T \to 0$ does not originate from electronic

| | relative orientation | $a$ | $n$ | $\Delta_0(0)/k_B T_c$ | $\sigma_{RMS}$ |
|---|---|---|---|---|---|
| Axial state | **I** ∥ **A** | $\pi^2$ | 2 | 1.923 | $9.85 \times 10^{-5}$ |
| | **I** ⊥ **A** | $7\pi^4/15$ | 4 | 0.969 | $1.72 \times 10^{-2}$ |
| Polar state | **I** ∥ **A** | $27\pi\zeta(3)/4$ | 3 | 1.284 | $5.82 \times 10^{-3}$ |
| | **I** ⊥ **A** | $3\pi \ln 2/2$ | 1 | 6.285 | $1.41 \times 10^{-2}$ |

Supplementary Table 1. (First four columns) Low temperature theoretical asymptotes of the normalized superfluid density $\rho_s(T) = 1 - a(k_B T/\Delta_0)^n$ with the predicted values for parameters $(a, n)$ for the various spin-triplet pairing states, and the relative direction between the symmetry axis of the gap **I** and the direction of the vector potential **A** (which is same as that of the current) [27, 39]. (Last two columns) The results of fitting the $\rho_s(T)$ data with the theoretical asymptotes of each scenario. Here, $\Delta_0(0)/k_B T_c$ was the single fitting parameter and $\sigma_{RMS}$ is the root-mean-square error of the fit.

entropy. Therefore, the residual $c/T$ due to the divergent term should not be attributed to bulk normal electronic states. Examination of other possible origins such as magnetic entropy or quantum critical fluctuations might be an interesting future research direction.

### SUPPLEMENTARY NOTE 8: DETAILS OF THE FITTING PARAMETER IN THE NORMALIZED SUPERFLUID DENSITY ANALYSIS

In this section, details of the normalized superfluid density $\rho_s(T)$ fit for the spin-triplet pairing states will be introduced for the reader's convenience. Note that one can also refer to Ref. [27, 39] for broader context and details. In general, if the symmetry axis of the gap is denoted as $\hat{\mathbf{I}}$, $\rho_s(T)$ for the case of $\hat{\mathbf{I}} \parallel \hat{\mathbf{A}}$ and $\hat{\mathbf{I}} \perp \hat{\mathbf{A}}$ can be calculated as,

$$\rho_s(T)_{(\parallel, \perp)} = 1 - 3n \int_0^1 d\cos\theta (\cos^2\theta, \frac{1}{2}\sin^2\theta) \int_{-\infty}^{+\infty} d\epsilon_k \left(-\frac{\partial f(E_k)}{\partial E_k}\right), \qquad (21)$$

where $E_k = \sqrt{\epsilon_k^2 + \Delta_k^2}$. In the case of the spin-triplet states, at low temperature the $\rho_s(T)$ has simpler theoretical asymptotes $\rho_s(T) = 1 - a(k_B T/\Delta_0)^n$. The values of the fitting parameter $\Delta_0(0)/k_B T_c$ from fitting the $\rho_s(T)$ data into these theoretical forms are shown in the main text Fig. 3(c) and Supplementary Table 1).

Here, the $\rho_s(T)$ data for $T \leq 0.26 T_c$ was fitted. The reason this temperature range is chosen is that the maximum gap magnitude $\Delta_0(T)$ is nearly constant for $T \leq 0.26 T_c$ [27].

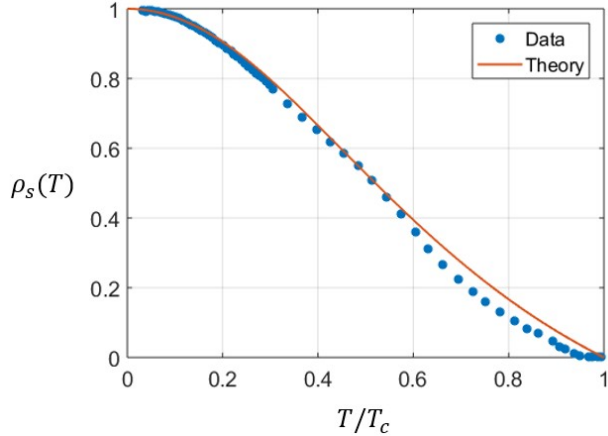

Supplementary Fig. 6. The $\rho_s(T)$ data of the UTe$_2$ sample for the full temperature range (0 to $T_c$) and a comparison to the theoretical estimation (Supplementary Equation (21)), using experimentally determined $\Delta_0(0)/k_B T_c$ and $\Delta c/c$ as discussed in the text.

One can estimate the temperature dependence of the gap by [39],

$$\Delta_0(T) = \Delta_0(0) \tanh\left(\frac{\pi k_B T_c}{\Delta_0(0)} \sqrt{\frac{2}{3} \frac{\Delta c}{c} \left(\frac{T_c}{T} - 1\right)}\right), \qquad (22)$$

where $\Delta c/c$ is the normalized jump in specific heat at $T_c$. $\Delta_0(T)$ shows only a 1.8 % decrease between 0 to 0.26 $T_c$. Therefore, for $T \leq 0.26 T_c$, $\Delta_0(T) \approx \Delta_0(0)$ so that one does not need to include $\Delta c/c$ as a fit parameter. The fewer the number of fitting parameters, the less arbitrary (and more accurate) the estimated parameter values are. This is the main reason why the superfluid density studies have been conducted effectively at $T \leq 0.26 T_c$ regime for the pairing state analysis [27].

To satisfy the reader's curiosity, if one plots the $\rho_s(T)$ data over the full temperature range (the same data displayed in Fig. 3(c) of the main text) with the theoretical estimation from Supplementary Equation (21) using $\Delta c/c = 1.18$ (obtained after subtracting the non-electronic contribution (from the data in Ref. [38])) and $\Delta_0(0)/k_B T_c = 1.923$ (obtained from the low temperature range $\rho_s(T)$ study), the data and the theory show good agreement (see Supplementary Fig. 6).

## SUPPLEMENTARY NOTE 9: INSULATING PROPERTIES OF URANIUM OXIDES

The UTe$_2$ sample has been prepared and handled in a nitrogen environment, such as a glove box, to exclude the potential effect of oxidizing the sample surface. Nevertheless, a thorough survey of the electrical properties of uranium oxides was conducted to address any concern that a metallic phase could exist on the surface. The early study of the electrical properties of the oxidized uranium was done on UO$_2$, which shows it is a semiconductor with a large band gap $\sim 2.1$ eV [40]. Another comprehensive study of the chemical constituents and their electronic properties of uranium oxide films shows that all of the common oxides of uranium (UO$_2$, U$_4$O$_9$, U$_3$O$_7$, U$_3$O$_8$, $\alpha$-UO$_3$, $\delta$-UO$_3$, $\gamma$-UO$_3$) are semiconductors with band gaps in the 1 to 3 eV range [41]. Similar results were recently obtained on epitaxial UO$_2$, $\alpha$-U$_3$O$_8$, and $\alpha$-UO$_3$ thin films [42]. At the temperatures of concern here ($< 3$ K, $k_B T < 0.26$ meV), they should all show an insulating response. This makes the oxidized surface, if there is any, less likely to produce the observed microwave dissipation and corresponding residual normal metallic behavior in the $T \to 0$ limit at 11 GHz ($\hbar\omega = 0.044$ meV).

## SUPPLEMENTARY NOTE 10: EFFECT OF N-GREASE ON THE SURFACE IMPEDANCE STUDY OF THE SAMPLE

In our study, we applied Apeizon N-grease between the sample and its thermal reservoir. The N-grease is the product name of a purified hydrocarbon low temperature adhesive that ensures thermal connection between the sample and the thermal reservoir at cryogenic temperatures, while being electrically insulating (https://apiezon.com/products/vacuum-greases/apiezon-n-grease/). Hence, it has been widely used for various microwave experiments in cryogenic conditions. To make sure that it didn't affect our microwave impedance measurement of the sample, we compared the $1/Q$ with and without a drop of N-grease present (which is the same amount we apply when we mount the sample) with the disk dielectric resonator setup. We only see a change in $1/Q$ of $\approx 10^{-6}$, which is near the resolution limit of our $1/Q$ measurement. This is marginal compared to 1/Q in the presence of the sample ($10^{-3} \sim 10^{-4}$). Therefore, we concluded the N-grease does not affect our results with the sample.

# SUPPLEMENTARY NOTE 11: EXAMINATION OF TOPOLOGICALLY TRIVIAL ORIGINS OF THE SURFACE NORMAL FLUID

From the theoretical point of view, it is proposed that a rough surface can suppress the order parameter of an anisotropic superconductor through diffusive scattering at the disordered surface [43] with thickness $d$ smaller than coherence length $\xi(0)$ [44]. The surface disorder provides a shorter mean free path $l_{\mathrm{mfp}}$ than the bulk, resulting in a large, dominant surface impurity scattering rate $\Gamma_{\mathrm{imp}}^{\mathrm{surf}}$. Once $\Gamma_{\mathrm{imp}}^{\mathrm{surf}}$ exceeds a certain critical value, the order parameter at such a surface can be suppressed. However, $\Gamma_{\mathrm{imp}}$ does not show noticeable temperature dependence as seen from Ti (430/490 GHz for below/above $T_c$ [20]). This implies that if a surface normal fluid is introduced by a dominant $\Gamma_{\mathrm{imp}}^{\mathrm{surf}}$ from the disorder of the rough surface, its contribution to $\sigma_1$ would not provide the more than factor of 2 enhancement ($\sigma_1(0)/\sigma_n > 2$) observed in our data.

Another topologically trivial origin for the surface normal fluid is the case of the clean surface with specular reflection. The specular reflection can break Cooper pairs in a superconductor whose order parameter changes sign for different directions of the momentum before and after the reflection. Theoretical studies expect this effect for $d$-wave superconductors to create a normal fluid region (Andreev bound state - ABS) at its surface with a thickness $L \sim \xi(0)$ [45]. The consequent experimental signatures of this surface ABS were observed as a zero-bias conductance peak in YBCO by point-contact spectroscopy [46].

However, the microwave conductivity studies of such systems [22, 47] do not show the monotonic increase in $\sigma_1$ below $T_c$ observed in our study (Fig. 2(b) in main text). Instead, they show a significant decrease in $\sigma_1$ as $T \to 0$ as seen from Fig. 9 of Ref. [47]). This is possibly because the thickness of the normal layer from ABS is very thin ($\xi(0) \approx 2$ nm [48]) compared to penetration depth $\lambda(0) \approx 155$ nm [49] which determines the probing depth of the electromagnetic response. In this case, the normal layer contribution to $\sigma_1$ can be marginal. Similarly, UTe$_2$ has $\xi(0) = \sqrt{\Phi_0/2\pi H_{c2}} = 3 \sim 7$ nm [13] and $\lambda(0) \approx 791$ nm, showing even a smaller ratio of $\xi(0)/\lambda(0)$ than YBCO. This fact makes the contribution of the normal fluid layer from the ABS to $\sigma_1$, even if the layer exists, really marginal. Note that the same thickness argument applies for the effect of a surface disordered layer on $\sigma_1$.

On the other hand, a naïve, but effective estimation of the thickness of the chiral surface state ($d_{css}$) can be given by the uncertainty principle $\Delta E \Delta t \sim \hbar/2$. As $\Delta t = d_{css}/v_F$ and

$\Delta E \sim \Delta_0(0)$, one obtains $d_{css} = \hbar v_F / 2\Delta_0(0)$. With $v_F \approx 3.9 \times 10^5$ m/s from ARPES study [26] and $\Delta_0 \approx 0.265$ meV from superfluid conductivity analysis in this work (and STM spectra [50]), $d_{css}$ is estimated to be 488 nm, which is comparable to $\lambda(0) \approx 791$ nm. The chiral surface normal state with the thickness $d_{css}$ comparable to $\lambda(0)$ could certainly provide significant contribution to the normal fluid response $\sigma_1(T)$. Note that the large residual normal fluid fraction $f_n(0) \approx 0.58$ calculated from the two-fluid model in Supplementary Note 6 (Although it is only a qualitative analysis) is similar to the ratio $d_{css}/\lambda(0) = 0.62$, providing more support to the scenario of the large normal fluid response from the chiral surface state. A qualitative argument explaining the monotonic increase of $\sigma_1(T)$ as $T \to 0$ is provided in the paragraph just before the conclusion of the main text.

## Supplementary References

[1] Booth, J. C., Wu, D. H. & Anlage, S. M. A broadband method for the measurement of the surface impedance of thin films at microwave frequencies. *Rev. Sci. Instrum.* **65**, 2082–2090 (1994). URL https://doi.org/10.1063/1.1144816.

[2] Scheffler, M., Dressel, M., Jourdan, M. & Adrian, H. Extremely slow Drude relaxation of correlated electrons. *Nature* **438**, 1135–1137 (2005). URL https://doi.org/10.1038/nature04232.

[3] Petersan, P. J. & Anlage, S. M. Measurement of resonant frequency and quality factor of microwave resonators: Comparison of methods. *J. Appl. Phys.* **84**, 3392–3402 (1998). URL https://doi.org/10.1063/1.368498.

[4] Kajfez, D. *Q factor* (Vector Fields, 1994).

[5] Matthaei, G. L. & Hey-Shipton, G. L. Concerning the use of high-temperature superconductivity in planar microwave filters. *IEEE Trans. Microw. Theory Tech* **42**, 1287–1294 (1994).

[6] Ghigo, G. *et al.* Effects of heavy-ion irradiation on the microwave surface impedance of $(Ba_{1-x}K_x)Fe_2As_2$ single crystals. *Supercond. Sci. Technol.* **31**, 34006 (2018). URL http://dx.doi.org/10.1088/1361-6668/aaa858.

[7] Huttema, W. A. *et al.* Apparatus for high-resolution microwave spectroscopy in strong magnetic fields. *Rev. Sci. Instrum.* **77**, 023901 (2006). URL https://doi.org/10.1063/1.2167127.

[8] Hein, M. A. *High-Temperature Superconductor Thin Films at Microwave Frequencies* (Springer, Heidelberg, 1999).

[9] Bonn, D. A. & Hardy, W. N. *Microwave Electrodynamics of High Temperature Superconductors* (Springer New York, New York, NY, 2007). URL https://doi.org/10.1007/978-0-387-68734-6_4.

[10] Zhang, X. *Microwave Flux-Flow Impedance Measurement of Type-II Superconductors* (Simon Fraser University, 2008). URL https://summit.sfu.ca/item/9887.

[11] Tobar, M. E., Krupka, J., Ivanov, E. N. & Woode, R. A. Anisotropic complex permittivity measurements of mono-crystalline rutile between 10 and 300 K. *Journal of Applied Physics* **83**, 1604–1609 (1998). URL https://doi.org/10.1063/1.366871.

[12] Bae, S. *et al.* Dielectric resonator method for determining gap symmetry of superconductors through anisotropic nonlinear meissner effect. *Rev. Sci. Instrum.* **90**, 043901 (2019). URL https://doi.org/10.1063/1.5090130.

[13] Ran, S. *et al.* Nearly ferromagnetic spin-triplet superconductivity. *Science* **365**, 684–687 (2019). URL https://science.sciencemag.org/content/365/6454/684.

[14] Sridhar, S. & Kennedy, W. L. Novel technique to measure the microwave response of high $T_c$ superconductors between 4.2 and 200 K. *Review of Scientific Instruments* **59**, 531–536 (1988). URL https://doi.org/10.1063/1.1139881.

[15] Rubin, D. L. *et al.* Observation of a narrow superconducting transition at 6 GHz in crystals of $YBa_2Cu_3O_7$. *Phys. Rev. B* **38**, 6538–6542 (1988). URL https://link.aps.org/doi/10.1103/PhysRevB.38.6538.

[16] Murphy, N. *Microwave frequency vortex dynamics of the heavy fermion superconductor $CeCoIn_5$*. Master's thesis (2010). URL http://summit.sfu.ca/item/12487.

[17] Ormeno, R. J., Sibley, A., Gough, C. E., Sebastian, S. & Fisher, I. R. Microwave Conductivity and Penetration Depth in the Heavy Fermion Superconductor $CeCoIn_5$. *Phys. Rev. Lett.* **88**, 047005 (2002). URL https://link.aps.org/doi/10.1103/PhysRevLett.88.047005.

[18] Ormeno, R. J. *et al.* Electrodynamic response of $Sr_2RuO_4$. *Phys. Rev. B* **74**, 092504 (2006). URL https://link.aps.org/doi/10.1103/PhysRevB.74.092504.

[19] Nelson, W. E. & Hoffman, A. R. *Measurements of the Temperatures and Magnetic Field Dependence of Electrical Resistivity and Thermal Conductivity in OFHC Copper*, 73–80 (Springer US, Boston, MA, 1976). URL https://doi.org/10.1007/978-1-4899-3751-3_13.

[20] Thiemann, M., Dressel, M. & Scheffler, M. Complete electrodynamics of a BCS superconductor with $\mu$eV energy scales: Microwave spectroscopy on titanium at mK temperatures. *Phys. Rev. B* **97**, 214516 (2018). URL https://link.aps.org/doi/10.1103/PhysRevB.97.214516.

[21] Quinlan, S. M., Scalapino, D. J. & Bulut, N. Superconducting quasiparticle lifetimes due to spin-fluctuation scattering. *Phys. Rev. B* **49**, 1470–1473 (1994). URL https://link.aps.org/doi/10.1103/PhysRevB.49.1470.

[22] Bonn, D. A. *et al.* Microwave determination of the quasiparticle scattering time in $YBa_2Cu_3O_{6.95}$. *Phys. Rev. B* **47**, 11314–11328 (1993). URL https://link.aps.org/doi/10.1103/PhysRevB.47.11314.

[23] Truncik, C. J. S. *et al.* Nodal quasiparticle dynamics in the heavy fermion superconductor CeCoIn5 revealed by precision microwave spectroscopy. *Nat. Commun.* **4**, 2477 (2013). URL https://doi.org/10.1038/ncomms3477.

[24] Sundar, S. *et al.* Coexistence of ferromagnetic fluctuations and superconductivity in the actinide superconductor UTe$_2$. *Phys. Rev. B* **100**, 140502 (2019). URL https://link.aps.org/doi/10.1103/PhysRevB.100.140502.

[25] Aoki, D. *et al.* Unconventional Superconductivity in Heavy Fermion UTe2. *J. Phys. Soc. Jpn.* **88**, 043702 (2019). URL https://doi.org/10.7566/JPSJ.88.043702.

[26] Miao, L. *et al.* Low Energy Band Structure and Symmetries of UTe$_2$ from Angle-Resolved Photoemission Spectroscopy. *Phys. Rev. Lett.* **124**, 076401 (2020). URL https://link.aps.org/doi/10.1103/PhysRevLett.124.076401.

[27] Prozorov, R. & Giannetta, R. W. Magnetic penetration depth in unconventional superconductors. *Supercond. Sci. Technol.* **19**, R41 (2006). URL http://stacks.iop.org/0953-2048/19/i=8/a=R01.

[28] Hirschfeld, P. J. & Goldenfeld, N. Effect of strong scattering on the low-temperature penetration depth of a d-wave superconductor. *Phys. Rev. B* **48**, 4219–4222 (1993). URL https://link.aps.org/doi/10.1103/PhysRevB.48.4219.

[29] Chia, E. E. M. *et al.* Nonlocality and strong coupling in the heavy fermion superconductor CeCoIn$_5$ : A penetration depth study. *Phys. Rev. B* **67**, 014527 (2003). URL https://link.aps.org/doi/10.1103/PhysRevB.67.014527.

[30] Radtke, R. J., Levin, K., Schüttler, H.-B. & Norman, M. R. Predictions for impurity-induced $T_c$ suppression in the high-temperature superconductors. *Phys. Rev. B* **48**, 653–656 (1993). URL https://link.aps.org/doi/10.1103/PhysRevB.48.653.

[31] Mackenzie, A. P. *et al.* Extremely Strong Dependence of Superconductivity on Disorder in Sr$_2$RuO$_4$. *Phys. Rev. Lett.* **80**, 161–164 (1998). URL https://link.aps.org/doi/10.1103/PhysRevLett.80.161.

[32] Nakamine, G. *et al.* Superconducting Properties of Heavy Fermion UTe2 Revealed by 125Te-nuclear Magnetic Resonance. *J. Phys. Soc. Jpn* **88**, 113703 (2019). URL https://doi.org/10.7566/JPSJ.88.113703.

[33] Tinkham, M. *Introduction to Superconductivity, 2nd edition* (Dover publications, Inc., 1996).

[34] Glover, R. E. & Tinkham, M. Conductivity of superconducting films for photon energies

between 0.3 and $40kT_c$. *Phys. Rev.* **108**, 243–256 (1957). URL https://link.aps.org/doi/10.1103/PhysRev.108.243.

[35] Klein, O., Nicol, E. J., Holczer, K. & Grüner, G. Conductivity coherence factors in the conventional superconductors Nb and Pb. *Phys. Rev. B* **50**, 6307–6316 (1994). URL https://link.aps.org/doi/10.1103/PhysRevB.50.6307.

[36] Culbertson, J. C., Newman, H. S. & Wilker, C. Optical probe of microwave current distributions in high temperature superconducting transmission lines. *J. Appl. Phys.* **84**, 2768–2787 (1998). URL http://dx.doi.org/10.1063/1.368390.

[37] Zhuravel, A. P., Ustinov, A. V., Harshavardhan, K. S. & Anlage, S. M. Influence of LaAlO3 surface topography on rf current distribution in superconducting microwave devices. *Applied Physics Letters* **81**, 4979–4981 (2002). URL http://dx.doi.org/10.1063/1.1530753.

[38] Metz, T. *et al.* Point-node gap structure of the spin-triplet superconductor UTe$_2$. *Phys. Rev. B* **100**, 220504 (2019). URL https://link.aps.org/doi/10.1103/PhysRevB.100.220504.

[39] Gross, F. *et al.* Anomalous temperature dependence of the magnetic field penetration depth in superconducting UBe$_{13}$. *Z. Phy. B* **64**, 175–188 (1986). URL https://doi.org/10.1007/BF01303700.

[40] Schoenes, J. Optical properties and electronic structure of UO$_2$. *J. Appl. Phys.* **49**, 1463–1465 (1978). URL https://doi.org/10.1063/1.324978.

[41] He, H., Andersson, D. A., Allred, D. D. & Rector, K. D. Determination of the insulation gap of uranium oxides by spectroscopic ellipsometry and density functional theory. *J. Phys. Chem. C* **117**, 16540–16551 (2013). URL https://doi.org/10.1021/jp401149m.

[42] Enriquez, E. *et al.* Structural and Optical Properties of Phase-Pure UO$_2$, $\alpha$-U$_3$O$_8$, and $\alpha$-UO$_3$ Epitaxial Thin Films Grown by Pulsed Laser Deposition. *ACS Appl. Mater. Interfaces* **12**, 35232–35241 (2020). URL https://doi.org/10.1021/acsami.0c08635.

[43] Barash, Y. S., Svidzinsky, A. A. & Burkhardt, H. Quasiparticle bound states and low-temperature peaks of the conductance of nis junctionsin d-wave superconductors. *Phys. Rev. B* **55**, 15282–15294 (1997). URL https://link.aps.org/doi/10.1103/PhysRevB.55.15282.

[44] Bakurskiy, S. V. *et al.* Local impedance on a rough surface of a chiral *p*-wave superconductor. *Phys. Rev. B* **98**, 134508 (2018). URL https://link.aps.org/doi/10.1103/PhysRevB.98.134508.

[45] Löfwander, T., Shumeiko, V. S. & Wendin, G. Andreev bound states in high-Tc supercon-

ducting junctions. *Supercond. Sci. Technol* **14**, R53–R77 (2001). URL https://doi.org/10.1088%2F0953-2048%2F14%2F5%2F201.

[46] Aprili, M., Badica, E. & Greene, L. H. Doppler Shift of the Andreev Bound States at the YBCO Surface. *Phys. Rev. Lett.* **83**, 4630–4633 (1999). URL https://link.aps.org/doi/10.1103/PhysRevLett.83.4630.

[47] Hosseini, A. *et al.* Microwave spectroscopy of thermally excited quasiparticles in YBa$_2$Cu$_3$O$_{6.99}$. *Phys. Rev. B* **60**, 1349–1359 (1999). URL https://link.aps.org/doi/10.1103/PhysRevB.60.1349.

[48] Varshney, D., Singh, R. K. & Shah, S. Coherence lengths and magnetic penetration depths in YBa$_2$Cu$_3$O$_7$ and YBa$_2$Cu$_4$O$_8$ superconductors. *J. Supercon.* **9** (1996-12).

[49] Tallon, J. L. *et al.* In-Plane Anisotropy of the Penetration Depth Due to Superconductivity on the Cu-O Chains in YBa$_2$Cu$_3$O$_{7-\delta}$, Y$_2$Ba$_4$Cu$_7$O$_{15-\delta}$, and YBa$_2$Cu$_4$O$_8$. *Phys. Rev. Lett.* **74**, 1008–1011 (1995). URL https://link.aps.org/doi/10.1103/PhysRevLett.74.1008.

[50] Jiao, L. *et al.* Chiral superconductivity in heavy-fermion metal UTe2. *Nature* **579**, 523–527 (2020). URL https://doi.org/10.1038/s41586-020-2122-2.

4242



\end{document}